\documentclass[preprint,3p,times,review]{elsarticle} 

\usepackage[utf8]{inputenc}
\usepackage{tikz}
\usepackage{pgfplots}
\usepackage{color}
\usepackage{lineno}
\linenumbers
\usepackage{float}      
\usepackage{caption}
\usepackage{subcaption}
\usepackage{xcolor}
\usetikzlibrary{patterns,shapes.geometric}
\usepackage{amsmath,amsthm,amsfonts,amssymb}
\usepackage{mathtools}
\usepackage{graphicx}
\usepackage{multirow}
\usepackage{soul}
\usepackage{textcomp}
\usepackage{cmll}
\pgfplotsset{compat=1.16}
\usepackage{ gensymb }

\usepackage{algorithm}
\usepackage{algpseudocode}
\usepackage{setspace}
\usepackage{cleveref }
\usepackage{url}

\newlength\myindent
\Crefname{figure}{Fig.}{Figs.}

\definecolor{Bleu_XBMA}{RGB}{5,112,176}
\definecolor{Jaune_XBMA}{RGB}{249,220,92}
\definecolor{green_Mod}{RGB}{27, 158, 119}
\definecolor{orange_Mod}{RGB}{217, 95, 2}
\definecolor{yellow_Mod}{RGB}{230, 171, 2}
\definecolor{grey_Mod}{RGB}{102, 102, 102}
\definecolor{XBMA_A}{RGB}{223, 101, 176}
\definecolor{XBMA_B}{RGB}{221, 28, 119}
\definecolor{darkorange}{RGB}{255,140,0}

\newcommand{\lsolid}  [1]   {\raisebox{2pt}{\tikz{\draw[#1,solid     ,line width=1pt](0,0)--(5mm,0);}}}
\newcommand{\ldash}   [1]   {\raisebox{2pt}{\tikz{\draw[#1,dashed    ,line width=1pt](0,0)--(5mm,0);}}}
\newcommand{\ldott}   [1]   {\raisebox{2pt}{\tikz{\draw[#1,dotted    ,line width=1pt](0,0)--(5mm,0);}}}
\newcommand{\ldashdot}[1]   {\raisebox{2pt}{\tikz{\draw[#1,dashdotted,line width=1pt](0,0)--(5mm,0);}}}
\DeclareRobustCommand\bp{\begin{tikzpicture} \draw[Bleu_XBMA,fill=Bleu_XBMA]  (0.25,0.5ex) circle (0.6ex) ; \draw[Bleu_XBMA,very thick] (0,0.5ex) -- (0.5,0.5ex) ; \end{tikzpicture} } 
\DeclareRobustCommand\yp{\begin{tikzpicture} \draw[Jaune_XBMA,fill=Jaune_XBMA]  (0.25,0.5ex) circle (0.6ex) ; \draw[Jaune_XBMA,very thick] (0,0.5ex) -- (0.5,0.5ex) ; \end{tikzpicture} }

\DeclareRobustCommand\bprectKE{\begin{tikzpicture} \draw[pattern color=, fill=green_Mod] (0,0) rectangle (0.25,0.20) ; \end{tikzpicture}}
\DeclareRobustCommand\bprectKO{\begin{tikzpicture} \filldraw[fill=orange_Mod, draw=orange_Mod] (0,0) rectangle (0.25,0.20) ; 
     \draw[pattern=crosshatch, pattern color=black,draw=black] (0,0) rectangle (0.25,0.20) ; \end{tikzpicture}}
\DeclareRobustCommand\bprectKL{\begin{tikzpicture} \filldraw[fill=yellow_Mod, draw=yellow_Mod] (0,0) rectangle (0.25,0.20) ; 
     \draw[pattern=grid, pattern color=black,draw=black] (0,0) rectangle (0.25,0.20) ; \end{tikzpicture}}
\DeclareRobustCommand\bprectSA{\begin{tikzpicture} \filldraw[fill=grey_Mod, draw=grey_Mod] (0,0) rectangle (0.25,0.20) ; 
     \draw[pattern=vertical lines, pattern color=black,draw=black] (0,0) rectangle (0.25,0.20) ; \end{tikzpicture}}
\DeclareRobustCommand\bprectXBMAA{\begin{tikzpicture} \filldraw[fill=XBMA_A, draw=XBMA_A] (0,0) rectangle (0.25,0.20) ; 
     \draw[pattern=north east lines, pattern color=black,draw=black] (0,0) rectangle (0.25,0.20) ; \end{tikzpicture}}
\DeclareRobustCommand\bprectXBMAB{\begin{tikzpicture} \filldraw[fill=XBMA_B, draw=XBMA_B] (0,0) rectangle (0.25,0.20) ; 
     \draw[pattern=north west lines, pattern color=black,draw=black] (0,0) rectangle (0.25,0.20) ; \end{tikzpicture}}

\newcommand{\cbox}[1]{\raisebox{\depth}{\fcolorbox{black}{#1}{\null}}}
\DeclareRobustCommand{\cbox}[1]{
  \begin{tikzpicture}
    \draw[fill=#1]  (0,0)  rectangle (0.25,0.20) ;
  \end{tikzpicture}
}

\makeatletter
\def\bstctlcite{\@ifnextchar[{\@bstctlcite}{\@bstctlcite[@auxout]}}
\def\@bstctlcite[#1]#2{\@bsphack
 \@for\@citeb:=#2\do{%
   \edef\@citeb{\expandafter\@firstofone\@citeb}%
   \if@filesw\immediate\write\csname #1\endcsname{\string\citation{\@citeb}}\fi}%
 \@esphack}
\makeatother

\newcommand{\ie}{\emph{i.e.}}
\newcommand{\q}{\mathbf{q}} 
\newcommand{\qmean}{\widetilde{\mathbf{q}}} 
\newcommand{\qfluc}{\mathbf{q}'} 
\newcommand{\Nav}{\mathcal{NS}} 
\newcommand{\D}{\mathcal{D}} 
\newcommand{\etal}{\emph{et al.}}
\newcommand{\QOIs}{\Delta} 
\newcommand{\QOIv}{\boldsymbol{\QOIs}} 
\newcommand{\NQ}{N_{\QOIv}} 
\newcommand{\Mouts}{\delta} 
\newcommand{\Moutv}{\boldsymbol{\Mouts}} 
\newcommand{\datasymb}{\Mouts} 
\newcommand{\datas}{\overline{\datasymb}} 
\newcommand{\datav}{\overline{\boldsymbol{\datasymb}}} 
\newcommand{\randobs}{\tilde{\datasymb}} 
\newcommand{\randobv}{\tilde{\boldsymbol{\datasymb}}} 
\newcommand{\ND}{N_{\randobv}} 
\newcommand{\modl}{M} 
\newcommand{\param}{\theta} 
\newcommand{\paramvof}[1]{\boldsymbol{\param}_{#1}} 
\newcommand{\RANS}{\mathcal{F}_m} 
\newcommand{\NM}{N_{\modl}} 
\newcommand{\Scen}{S} 
\newcommand{\x}{\mathbf{x}} 

\newcommand{\mean}[1]{E\left[#1\right]} 
\newcommand{\var}[1]{Var\left[#1\right]} 
\newcommand{\M}{\mathcal{M}} 
\newcommand{\DPS}{\displaystyle}

\newcommand{\keps}{k-\varepsilon}
\newcommand{\truevals}{\hat{\Mouts}} 
\newcommand{\X}{\mathbf{X}} 
\newcommand{\Etab}{\boldsymbol{\mathrm{H}}} 
\newcommand{\Reyn}{R_e} 
\newcommand{\kl}{k-L}
\newcommand{\kkl}{k-kL}
\newcommand{\cf}{g} 
\newcommand{\hypar}{\sigma} 
\newcommand{\QOI}{\Delta}
\newcommand{\expcomp}{\eta} 
\newcommand{\etav}{\boldsymbol{\expcomp}} 
\newcommand{\C}{\mathbf{C}}

\newcommand{\Ps}{\mathbf{P}}
\newcommand{\NP}{N_{\Ps}}
\newcommand{\coff}{C} 
\newcommand{\St}{\mathbf{S}} 
\newcommand{\Ot}{\boldsymbol{\Omega}} 
\newcommand{\vel}{U} 
\newcommand{\pres}{P} 
\newcommand{\derp}[2]{\dfrac{\partial #1}{\partial #2}}

\DeclareUnicodeCharacter{2212}{-}

\title{Space-dependent turbulence model aggregation using machine learning}

\author{\textbf{\normalsize M. de Zordo-Banliat$^{(1,2)}$, G. Dergham$^{(1)}$, X. Merle$^{(2)}$, P. Cinnella$^{(3)}$}
\\{\normalsize\itshape $^\text{(1)}$ Safran  Tech, Digital Sciences \&  Technologies Department, Rue
  des Jeunes Bois, Châteaufort, 78114 Magny-Les-Hameaux, France } 
\\{\normalsize\itshape  $^\text{(2)}$  DynFluid Laboratory  -  Arts  et  M\'etiers ParisTech  -  151
  boulevard de l'H\^opital, 75013 Paris, France} 
\\{\normalsize\itshape $^\text{(3)}$  Institut Jean Le  Rond D'Alembert  - Sorbonne Univerit\'e  - 4
  Place Jussieu, 75005 Paris, France} 
\\{\normalsize\itshape  gregory.dergham@safrangroup.com }}

\begin{document}
\nolinenumbers
\begin{abstract} 
 Computational models of fluid flows based on the Reynolds-averaged Navier--Stokes  (RANS) equations supplemented with a turbulence model
  are the golden standard in engineering applications.  A plethora of turbulence models and related variants exist, none of which is fully reliable outside the range
  of flow configurations for which they have been calibrated. Thus, the choice of a suitable turbulence closure largely relies on subjective expert judgement and engineering know-how.
  In this  article, we propose a data-driven methodology for combining the solutions of a set of competing turbulence models.
  The individual model predictions are linearly combined for providing an ensemble solution accompanied by estimates of
  predictive uncertainty due to the turbulence model choice.
   First, for a set of training flow configurations we assign to component models high weights in the regions where they best perform, and vice versa, by introducing a measure of  distance between high-fidelity data and individual model predictions.   The model weights are then mapped into a space of features, representative of local flow physics, and regressed by a Random Forests (RF) algorithm.  The RF regressor is finally employed to infer spatial distributions of the model weights for unseen configurations. 
 Predictions of new cases are constructed as a convex linear combination of the underlying models
solutions, while the between model variance provides information about regions of high model uncertainty.
  The  method is demonstrated for a class of flows through the compressor
  cascade NACA65 V103 at $\Reyn\simeq 3\times 10^5$. The results show that  the aggregated solution outperforms the accuracy of individual models for the quantity used to inform the RF regressor, and performs well for other quantities well-correlated to the preceding one. The estimated uncertainty intervals are generally consistent with the target high-fidelity data.
 The present approach then represents a viable methodology for a more objective selection and combination of alternative turbulence models in configurations of interest for engineering practice.
\end{abstract}


\maketitle

\section{Introduction}
Scientific modeling  is the process of  describing the physical reality  via mathematical equations,
based on a set of simplifying assumptions and on experimental observation of a system.  The modeling
hypotheses determine the  mathematical structure of the  model, called a model class,  and limit the
validity of its  predictions.  A specific model is  then singled-out from the class  by specifying a
set of  closure parameters, generally  adjusted to fit experimental  observation. Such a  process is
called \emph{calibration}.  Due to the underlying  modeling assumptions
and to observation errors corrupting experimental data, hence closure parameters,
predictions based on a model are affected by  uncertainties on both the model form (called epistemic
uncertainty) and on the closure parameters (parametric uncertainty) \cite{oden_computer_2010}.

In the present work we focus on Computational  Fluid Dynamics (CFD) models, i.e. computer models for
solving the governing equations for fluid flows:

\begin{equation}\label{NavierStokes}
  \Nav[\q(\x,t);\Scen]=0
\end{equation}
where $\Nav$  is the Navier-Stokes partial  differential operator, $\q(\x,t)$ is  the state variable
vector (e.g. pressure,  velocity,...), $\x$ is a vector of  coordinates in the geometrical
space, $t$ is the time  and $\Scen$ is a set of parameters determining  the flow conditions, such as
the geometry and  the initial and boundary  conditions, which we call a  flow \emph{scenario}.  Very
broad model classes, better  called modeling "levels", can be identified  according to the inclusion
or not  of some physical effects.   We distinguish for  instance inviscid from viscous  models.  For
viscous  flows characterized  by  high values  of  the Reynolds  number ($Re=U  L/\nu$,  with $U$  a
characteristic  velocity scale,  $L$ a  length scale,  and  $\nu$ a  reference value  for the  fluid
kinematic viscosity) several  modeling levels can be  identified, according to the  strategy used to
account for turbulent  motions.  Direct Numerical Simulations (DNS) solve  for all turbulent motions
by seeking the solution of the discretised form of \eqref{NavierStokes}:
\begin{equation}
  \D\left[\Nav[\q(\x,t);\Scen]\right]=0,
\end{equation}
with $\D$  a discretisation operator on a fine-grained space and time mesh.  The computational cost  of DNS models  scales as $Re^{11/4}$  for DNS
\cite{pope2000turbulent},  making  them  prohibitively  expensive  for  the
high-Reynolds  number  flows  of  interest.   
On the other  hand, by
applying a  coarse-grained operator to  Eq. \eqref{NavierStokes}, state variables can  be split
into a coarse-grained  part $\qmean$ and an unresolved $\qfluc$ part:  $\q=\qmean+\qfluc$.  
Large-Eddy
Simulations (LES), consisting in filtering the small scales and resolving the large, energetic ones, require a number of  grid  points of  order  of $Re^{13/7}$  for resolving
wall-bounded flows  \cite{choi_requirements_2012}, which strongly limits their routine use for practical engineering problems.
As  a result,  engineering  design  mostly  relies  on
lower-fidelity models such as the Reynolds-averaged  Navier–Stokes (RANS) equations.  The latter use
a similar decomposition to LES whereby the filter is replaced by a statistical averaging operator:
\begin{equation}
  \widetilde{\Nav}[\q(\x,t);\Scen]=0\quad\Rightarrow\quad
  \Nav[\qmean;\Scen] + \nabla\cdot\mathcal{F}(\qmean,\qfluc)=0
\end{equation}
Thus the RANS  equations solve only for mean flow  motions.  $\mathcal{F}(\qmean,\qfluc)$ represents
the contribution of turbulent  flow scales to the transport of momentum and  energy and is accounted
for by supplementing the RANS equations with additional constitutive relations. The latter, called a
turbulence model, express the so-called Reynolds stress  tensor (a measure of turbulent transport of
momentum) as a function of mean field variables:
\begin{equation}\label{RANSequ}
  \Nav[\qmean_m;\Scen] + \nabla\cdot\RANS(\qmean_m;\paramvof{\RANS})=0
\end{equation}
where  $\qmean_m$  is  an  approximation  of   $\qmean$  under  the  chosen  turbulence  model,  and
$\paramvof{\RANS}$ is a vector of parameters associated with the turbulence model $\RANS$.  The RANS
equations \eqref{RANSequ} are then numerically solved by applying some discretisation operator $\D$:
\begin{equation}\label{discrete_RANS}
  \D \left[\Nav[\qmean_m;\Scen] + \nabla\cdot\RANS(\qmean_m;\paramvof{\RANS})\right]=0
\end{equation}
Only     steady     solutions     are     sought     in     this     work,     and     we     denote
$\qmean_m^{\dagger}=\qmean_m^{\dagger}(\x)$  the  numerical  solution   of  the  discrete  equations
\eqref{discrete_RANS}. For a  Quantity of Interest (QoI) $\Mouts$, post-processed  from the computed
state vector $\qmean_m^{\dagger}$, the whole process may be written in short as
\begin{equation}\label{Mcfd}
  \Mouts=\modl(\x;\Scen,\RANS,\paramvof{\RANS})
\end{equation}
where $\modl$ is  the post-processed CFD model  output, which depends on the  geometrical space, the
operating  conditions  $\Scen$,  and  the  turbulence  model.  For  brevity,  $\Scen$,  $\RANS$  and
$\paramvof{\RANS}$ are omitted when irrelevant.

Despite  the plethora  of turbulence  models proposed  in the  literature for  more than  a century,
determining  a turbulent  closure  universally valid  for  any  kind of  flow  remains a  formidable
challenge  (see  \cite{wilcox2006turbulence,spalart_philosophies_2015}   for  overviews).   Existing
turbulence  models can  be classified  according to  their  levels of  complexity, and  a number  of
variants and corrections  exist for different flow features such  as pressure gradients, separation,
vortices,  rotations,   shock  waves,  etc.   The   reader  is  referred  to   the  NASA  repository
\url{https://turbmodels.larc.nasa.gov}, for  the description of some  widely-used turbulence models.
Assessment against  academic and  industrial applications  shows that the  choice of  the turbulence
model and of the associated closure parameters $\paramvof{\RANS}$ may cause large variability in CFD
predictions, which in turn may be critical for decision making \cite{XiaoCinnella2019}.

Early attempts  to quantify uncertainties in  RANS models are  due to Cheung et  al.  \cite{Cheung},
Oliver  and Moser  \cite{OliverMoser_2011},  Emory  et al.   \cite{Emory_2013}  and  Edeling et  al.
\cite{edeling2014predictive}.    Such  studies   treat  turbulence   modeling  uncertainties   in  a
probabilistic framework:  instead of producing a  single deterministic prediction associated  with a
model form  and a  set of parameters,  they try  to estimate the  probability distribution  of model
outputs, conditioned on some random inputs.  The analysis is conducted either by perturbing directly
the     Reynolds    stress     anisotropy     tensor    computed     with     a    baseline     LEVM
\cite{Emory_2013,gorle_framework_2013,thompson_eigenvector_2019} or by treating the turbulence model
closure    parameters   as    random   variables    with   associated    probability   distributions
\cite{platteeuw_uncertainty_2008,Cheung,Edeling2014a,margheri2014epistemic}.    While    the   first
approach accounts for model-form uncertainties, the second  does not. In turn, the first approach is
intrusive in the sense  that its implementation involves modifications of the  RANS solver, while in
the second one the stochastic parameters can be just  fed as inputs to the CFD solver and propagated
by using a suitable (non intrusive) uncertainty quantification method.

An attractive  and non-intrusive approach for  quantifying model-form uncertainty is  represented by
multi-model,  or  ensemble,  statistical  methods.   For  instance,  \cite{Poroseva_2006}  used  the
Demster-Shafer  evidence theory  along with  the $k-\epsilon$  and $k-\omega$  turbulence models  to
quantify the variability of CFD  simulations. More recently, \cite{edeling2014predictive} explored a
Bayesian framework, namely,  Bayesian Model-Scenario Averaging (BMSA), to calibrate  and combine the
predictions obtained from  a set of competing  baseline LEVM models calibrated on  various data sets
(scenarios).  BMSA has been successfully applied to  provide stochastic predictions for a variety of
flows,    including     3D    wings     \cite{edeling2018bayesian}    and     compressor    cascades
\cite{MZB_2020,MZB_2022}.

BMSA, and Bayesian Model  Averaging (BMA) \cite{draper1995assessment,hoeting1999bayesian} from which
it originates, may be interpreted as stochastic  variants of the Model Aggregation framework (Stoltz
\cite{stoltz_agregation_2010},  Deswarte  \etal{} \cite{deswarte_sequential_2019},  Devaine  \etal{}
\cite{devaine_forecasting_2013}),  also  referred-to as  Sequential  Model  Aggregation (SMA).   The
latter belongs  to a wider class  of methods, denominated Multiplicative  Weight Updating Algorithms
\cite{blum_empirical_1997}, first  introduced in the late  80's \cite{littlestone_learning_1988} and
further        developed        in       \cite{desantis_learning_1988,        cesa-bianchi_how_1993,
  littlestone_weighted_1994}.   Such methods  aim at  combining multiple  predictions stemming  from
various models --also termed experts or forecasters--  to provide a global, enhanced, solution --- the
'wisdom of crowd' paradigm,  well-known in Machine Learning.  In SMA, weights  are computed by using
the Exponentially Weighted  Average (EWA) \cite{deswarte_sequential_2019}, which  may be interpreted
as a loss  function.  In such an approach  the parameters intrinsic to the individual  models may be
updated or  not during the training  process.  In BMA, the  predictions of the competing  models are
weighted  by  posterior  model  probabilities,  computed  through  the  Bayes'  theorem  of  inverse
probability.  Model  Aggregation is generally not  applied to space-dependent predictions  while BMA
has  been.   In  such   case,  however, the  same BMA  weights  are  assigned throughout  the  spatial  domain.
Spatially-constant weighting of different RANS-model solutions is not optimal, since prior knowledge
about RANS indicates that  the accuracy of models may vary according to  the local flow physics.  In
principle, one would like to assign higher weights to the best-performing models in each region.
  
Other   classes  of   ensemble  methods   allow  space-varying   weights  \cite{Breiman_1996Bagging,
  Breiman_2001RF, schapire_boosting_1998}.  In Mixture-of-local-Experts \cite{jacobs_adaptive_1991},
also  referred-to  as Mixture-of-Experts  \cite{yuksel_twenty_2012}  or  Mixture Models,  the  input
feature space  (covariate space) is softly  split into partitions where  the locally best-performing
models are assigned  higher weights.  The soft partitioning is  accomplished through parametric gate
functions, or a  network of hierarchical gate  functions \cite{JordanJacobs_1994expertregions}, that
rank the  model outputs with probabilities.   For promoting the  best models in each  partition, the
softmax  function \cite{bridle_probabilistic_1990,gao_properties_2018}  --- a smoothed  version of  the
winner-takes-all model  --- is  employed to  build the probabilities.   Parameters associated  with the
component  models and  with the  gate functions  are trained  simultaneously though  the Expectation
Maximisation      algorithm       \cite{dempster_maximum_1977},      or       improved      versions
\cite{baldacchino_variational_2016}.   Although  the  Mixture-of-Experts method  originates  from  a
stochastic formulation, the model output is  ultimately a deterministic convex linear combination of
individual expert outputs. Additionally, Mixture-of-Experts tends  to promote a single best model in
every soft  partition, thus  accounting for  the spatial variation  of the  best model  but strongly
neglecting the uncertainty in model choice.
%
  
In  the  attempt  of  combining  the  best  features  of  BMA  and  Mixture-of-Experts,  Yu  \etal{}
\cite{Yu_2013} proposed an improved version of the BMA, called "Clustered Bayesian Averaging" (CBA),
allowing for space-varying weighting  of the component models in different  regions of the covariate
space.  For  that purpose,  they computed  so-called "local Bayes  factors", which  were input  to a
clustering  algorithm  determining  a  spatial  partition.   Specifically,  the  Classification  And
Regression Trees  (CART) algorithm was employed  to identify the  clusters and to regress  the local
model  Bayes factors  as functions  of space.   At the  same time,  part of  the data  are used  for
iteratively updating  the parameters of  each component model.   Predictions of a  new configuration
through the CBA model  are finally constructed by estimating the  posterior model probabilities from
the regressed local Bayes  factors, and by assigning them as weights of  the model mixture.  The CBA
algorithm has been applied successfully  to hydrology \cite{Rahman_2020CBAHydro} and solid mechanics
\cite{Abdallah_2020CBAMeca}, but no extensions to CFD problems have been considered up to date.  CBA
represents an attractive approach for estimating  and improving turbulence modeling uncertainties in
CFD  because it  provides local  estimates of  posterior model  probabilities, used  for aggregating
component models into a "hypermodel" (the  model mixture) with improved predictive capabilities than
the mixture  components. The procedure  assigns weights  to the models,  as function of  local model
performance in each flow subregion, while adjusting their parameters.  The iterative update of model
parameters from data  improves accuracy via calibration,  but it also represents a  critical step of
the CBA algorithm, due to the huge number  of model evaluations, i.e. costly CFD solves, required in
the process.

In  the present  paper, we  propose and  validate a  novel space-dependent  Model Aggregation  (XMA)
algorithm to generate mixtures of RANS  solutions obtained with different turbulence models.  Unlike
the           BMSA           approach           used           in           previous           works
\cite{edeling2014predictive,edeling2018bayesian,MZB_2020,MZB_2022},  XMA  enables  locally  variable
weights.  The algorithm is  inspired from both CBA and SMA, redesigned  for costly CFD applications.
More precisely,  the original CBA  algorithm is simplified  to account for  the high
computational cost of CFD solves, and no model calibration is performed, so that local Bayes factors are no longer available. 
In the proposed methodology, we use instead an EWA loss function inspired from SMA to assign local model weights. The latter do not depend directly on geometrical
coordinates  in  the  physical space  but  on  a  set  of  well-chosen flow  features,  which  eases
generalization  of the  learned XMA  model to  different flows.   The XMA  algorithm is  trained and
applied to the  prediction of flows through  the compressor cascade NACA$65$  V$103$, showing better
accuracy than the individual RANS models in the mixture for all flow quantities of interest.

%
%
%

The paper is organized  as follows.  In Section \ref{section_XBMA}, we  present the XMA methodology.
Section  \ref{section_Refdata} provides  information  about the  RANS  models in  use  and the  flow
configuration and reference  data used in the numerical  experiments.  Section \ref{section_Results}
reports a detailed assessment of the proposed method for both interpolation and extrapolation cases,
and  for  different  training datasets.   Finally,  the  main  findings  are summarized  in  Section
\ref{chapXBMA_conclusion}, alongside perspectives for future developments.

\section{Space-dependent Model Aggregation (XMA)}\label{section_XBMA}
Be $\truevals(\x)$ the true value of a spatially-varying flow
quantity.
A modeled counterpart $\Mouts$ for $\truevals$ is obtained as the output of a CFD model of the form
\eqref{Mcfd}, which reads:
\begin{equation*}
  \Mouts=\modl(\x;\Scen,\RANS,\paramvof{\RANS}).
\end{equation*}
We assume that all inputs $\Scen$ to the CFD code (geometry, boundary conditions, etc) are perfectly
known  and numerical  errors are  negligibly  small, so  that  all deviations  between the  truthful
quantity $\truevals$ and the  model output $\Mouts$ are due to the turbulence  model $\RANS$ and its
closure parameters $\paramvof{\RANS}$.

In the  aim of accounting  to some extent for  model-form uncertainties, while  improving prediction
accuracy,  we do  not predict  $\truevals$ using  a single,  uncertain model.   Instead, we  adopt a
multi-model ensemble approach and  we construct a convex linear combination of  a set of alternative
RANS models,  or component  models, by means  of weighting functions  that depend  on a set  of flow
features (introduced in Section \ref{section_InputFeat}).
More precisely, we consider a discrete set of $\NM$ models
$$\M=\{\modl_1,...,\modl_m,...,\modl_{\NM}\}$$  
corresponding to $\NM$ CFD solves of the same problem based on different turbulence models:
$$\modl_m=\modl(\x;\Scen,\RANS,\paramvof{\RANS}),\quad m=1,...,\NM$$
The space-dependent model aggregation (XMA) then takes the form:
\begin{equation}\label{xma}
\sum_{m=1}^{\NM} w_m \modl_m
\end{equation}
where $w_m=w_m(\x)$  is a space-dependent  weighting function  associated with the  $m$-th component
model, subject to:
$$0\le w_m(\x) \le 1,\quad\text{and}\quad
\sum_{m=1}^{\NM} w_m(\x)=1\quad \forall \x$$
The closure  parameters $\paramvof{\RANS}$  associated to  $\RANS$ are also  uncertain and  could be
calibrated  from  data  (see  \cite{edeling2014bayesian,edeling2014predictive}).   However,  such  a
process generally requires a significant number of CFD solves for finding the best-fit values of the
parameters with  respect to  a set of  observed data. The  cost can  be alleviated, e.g.  by using
surrogate  models \cite{MZB_2020},  but it  remains high  for complex  3D configurations.   For that
reason, in the present study we do not  attempt to update the turbulence model parameters, which are
kept fixed to their  nominal values for each component model.  Since the  closure parameters are now
assigned once and for all for each turbulence model, we simplify the notation as follows:
$$\modl_m=\modl(\x;\Scen,\RANS),\quad m=1,...,\NM$$

The next  step of XMA  consists in inferring  the weighting functions of  the model mixture.  
\newline

Be  $\datav=(\datas_1,...,\datas_d,...,\datas_{\ND})^T$  a  vector  of  $\ND$  observations  of  the
quantity $\truevals$ at various spatial locations $\x_d\in\X$, $\X=\left\{\x_d\right\}_{d=1}^{\ND}$,
and possibly for various flow scenarios $\Scen$.   To enable the use of heterogenous observations in
$\datav$ (i.e.   corresponding to different flow  properties and flow conditions),  each subgroup of
data  is  assumed   to  be  standardized  to   a  distribution  of  zero  mean   and  unit  standard
deviation.\newline

We note  $\Moutv^{(m)}=(\Mouts_1^{(m)},...,\Mouts_d^{(m)},...,\Mouts_{\ND}^{(m)})^T$ the predictions of  a component
model $\modl_m$  at the observation  locations, where  $\Mouts_d^{(m)}=M_m(\x_d)$ (we omitted  the other
arguments for brevity).  Spatial  coordinates are specific to a given flow  configuration and do not
possess the due invariance properties for ensuring  model generalization.  For that reason, we chose
instead   to  transform   the  spatial   fields  into   a  well-chosen   space  of   \emph{features}
$\etav=\etav(\x)$,  i.e. flow  properties  representative  of the  local  flow  physics.  A  similar
approach is adopted, e.g., in  \cite{parish2016paradigm} to generalize spatially-dependent corrective
fields for the turbulence model transport equations.\newline

The mixture  weighting functions are then  learned in the  feature space instead of  the geometrical
space:
$$w_m=w_m(\etav)$$ 
by     transforming    the     training     dataset    $\X=\left\{\x_d\right\}_{d=1}^{\ND}$     into
$\Etab=\left\{\etav_d\right\}_{d=1}^{\ND}$.   In  practice,  the  exact features  are  unknown,  and
features estimated  from CFD depend  on the  turbulence model in  use.  Various strategies  are possible, such  as averaging  features estimated  from various  models.
Hereafter  we chose to  make each weighting  function dependent  on features
  estimated from the corresponding component model, i.e.
$$w_m=w_m(\etav^{(m)})$$

To compute the weighting functions  at points of the feature space not included  in the data set, we
use a supervised machine learning procedure. The weighting criteria, the machine learning regressor,
and  the  definition of  the  feature  space  are  discussed in  Sections  \eqref{weighting_models},
\eqref{regressor}, and \eqref{section_InputFeat}, respectively.  

\subsection{Weighting criteria}\label{weighting_models}
Several weighting functions are  available in the literature.  In this work,  we adapt the weighting
function from the  Exponentially Weighted Average (EWA) predictor, initially  introduced by Deswarte
\etal{}  \cite{deswarte_sequential_2019} for  SMA, by  introducing local  dependence on  the feature
vectors:
\begin{equation}\label{weight_equ}
  w_m(\Mouts^{(m)};\etav_d^{(m)},\datas_d,\hypar)                                              =
  \frac{g_m(\Mouts^{(m)};\etav_d^{(m)},\datas_d,\hypar)}{\DPS\sum_{j=1}^{\NM}
    g_j(\Mouts^{(j)};\etav_d^{(j)},\datas_d,\hypar)}, \quad m\in\{1,\cdots,\NM \} 
\end{equation}
where $g_m$ is a cost function defined by
\begin{equation}
  g_m         =        \exp         \left(-\frac{1}{2}        \frac{(\Mouts^{(m)}(\etav_d^{(m)})
      -\datas_d)^2}{\hypar^2}\right). 
\end{equation}
$g_m$ is reminiscent of a Gaussian likelihood function  used in Bayesian approaches, which amplifies/damps 
the discrepancies between the output of the $m$-th model and the data.  The
cost     function    equals     1    when     the     model    perfectly     matches    the     data
($\Mouts^{(m)}(\etav_d^{(m)})=\datas_d$)  and  it  tends  to 0  for  very  large  discrepancies,
respectively.    The   squared   exponential   ensures    a   smooth   variation   of   $g_m$   with
$\Mouts^{(m)}(\etav_d^{(m)})-\datas_d$.  The  parameter $\hypar$  is the  learning rate  of EWA,
controlling  how fast  the departure  of  the model  prediction  $\Mouts^{(m)}$ from  the observed  data
$\datas$ is  penalized by the  cost function $g_m$:  when $\hypar\rightarrow\infty$, the models are
assigned uniform weights; when $\hypar\rightarrow 0$, the worst-performing models get weights closer
to 0,  while the weight  of the best-performing  model gets closer to  1.  In other  terms, $\sigma$
controls model selection.  As in \cite{deswarte_sequential_2019}, the optimal value of the parameter
$\hypar$ is sought by  a grid search procedure.  Precisely, we compute the  Mean Squared Error (MSE)
between the final prediction and the observables for every point in the data set, such that :
\begin{equation}
  \hypar_{\text{opt}}    =     \text{arg}\min_{\hypar\in\Sigma}    \frac{1}{\ND}    \sum_{d=1}^{\ND}
\left(\datas_d            -            \sum_{m=1}^{\NM}            w_m(\etav_d^{(m)};\hypar)
  \Mouts^{(m)}(\etav_d^{(m)})\right)^2 
\end{equation}
where $\Sigma$ is  a grid of prescribed values  for $\hypar$.  We verified that  $\hypar$ has little
influence   on    the   model   accuracy,   provided    the   order   of   magnitude    is   correct
\cite{deswarte_sequential_2019}.



Finally, the  XMA prediction  $\Mouts$ of  $\truevals$ at point  $\etav_*$ of  the feature  space is
obtained as
\begin{equation}\label{prediction}
  \Mouts(\etav_*) = \sum_{m=1}^{\NM} w_m(\etav_*^{(m)}) \Mouts^m(\etav_*^{(m)})
\end{equation}
where we omitted the dependency on $\Mouts^{(m)}$, $\datas_d$ and $\hypar$ in $w_m$ for brevity.
Note that  the prediction  is dependent on  $\etav_*$, which  stands for the  concatenation of  feature vectors
$\etav_*^{(m)}$ for each component model. The subscript ${}_*$ 
refers to unobserved locations  in geometrical  space for the  training scenarios,  but also  to any
location of  a new (unseen)  scenario to  predict.  Since the  $g_m(\etav^{(m)})$ are known  only at
observation  points  $\etav_d^{(m)}$, they  are  regressed  across  the  feature space for each model $m$ to  obtain
estimates of the weights at new locations  $\etav_*$.  For that aim, a supervised regression method,
presented in Section \eqref{regressor} is employed. \newline

The  predictions $\Mouts^{(m)}$  given by  competing models  may be  interpreted as  the $\NM$  possible
outcomes of a discrete random variable  $\randobs$ whose probability mass function (pmf) corresponds
to the weights $w_m$.  Thereby, Eq.  \eqref{prediction}  can be interpreted as the expected value of
$\randobs$:
\begin{equation}\label{equ_esp_XBMA}
  \Mouts(\etav_*)=\mean{\randobs(\etav_*)}=\sum_{m=1}^{\NM}                               w_m(\etav_*^{(m)})
  \Mouts^{(m)}(\etav_*^{(m)}) 
\end{equation}
The aggregated prediction,  given by \eqref{equ_esp_XBMA}, gives more weight  to the best performing
models, and less  weight to the worst  performing ones.  As a  result, it is expected  be more
accurate than the individual models. Moreover, promoting the best  models is done locally in space through
the use  of the  flow features,  so that  each component  model participates  primarily at  the most
relevant locations. \newline 

Similarly, the predictive variance can be written as
\begin{equation}\label{equ_var_XBMA}
  \var{\randobs(\etav_*)}=\sum_{m=1}^{\NM}             w_m(\etav_*^{(m)})             \left(
    \Mouts^{(m)}(\etav_*^{(m)}) - \mean{\randobs(\etav_*^{(m)})}\right)^2 
\end{equation}
The variance \eqref{equ_var_XBMA}  must be understood as  an indicator of the  consensus among the
models: small variances result from a strong agreement of individual model predictions while large variances
reveal a divergence. Furthermore,  as the weights are better informed, they get closer  to 1 for the best
models and to 0  for the worst models, and the  variance decreases. This indicates that the  uncertainty about
model choice has been reduced. 
\newline

In practice, we may wish to use XMA  for predicting different QoI than the observed vector $\datav$,
either by direct extraction from the model  or through post-processing of the solution.  Examples of
such QoI  are given  by flow  properties at  given locations  in the  flow, like  velocity profiles,
pressure or skin friction distributions.  We denote $\QOIv=(\QOIs_1 ,..., \QOIs_{\NQ})^T$ the vector
of such  unobserved QoI.  Such outputs  also depend on the  flow scenario $\Scen$, the  equations of
motion and the turbulence model, \ie:
\begin{equation}\label{mod_deter2}
  \QOIs = \modl_\Delta(\x;\Scen,\RANS,\paramvof{\RANS})     
\end{equation}
where $\modl_\Delta$ stands for a different postprocessed output of the same flow solver than $\modl$.
The  model aggregation \eqref{equ_esp_XBMA}  and variance equation \eqref{equ_var_XBMA}  can be
transposed to the quantity $\QOI$. Since no data is available to inform the weights for $\QOIs$,
we use the same weights as those learned for $\randobs$:
\begin{equation}\label{equ_esp_QOI}
\QOI(\etav_*)=  \mean{\widetilde{\QOI}(\etav_*)}=\sum_{m=1}^{\NM} w_m(\etav_*^{(m)}) \QOI^{(m)}(\etav_*^{(m)}) 
\end{equation}
and
\begin{equation}\label{equ_var_QOI}
  \var{\widetilde{\QOI}(\etav_*)}=\sum_{m=1}^{\NM}               w_m(\etav_*^{(m)})               \left(
    \QOI^{(m)}(\etav_*^{(m)}) - \mean{\QOI(\etav_*)}\right)^2 
\end{equation}
where $\QOI^{(m)}$  is the $\QOI$  output for the $m$-th model. Using the same weights for $\Mouts$ and $\QOI$ is reasonable as  long as  the deviations from  observed to predicted  quantities, of
$\randobs$ are correlated to those of $\QOI$.  The  latter is a desirable property meaning that both
$\randobs$ and $\QOI$ are  well predicted on the same locations.  In this  case, the space dependent
weights in \eqref{equ_esp_QOI} and \eqref{equ_var_QOI} will promote models where both $\randobs$ and
$\QOI$  are accurately  predicted.   As a  result,  we can  also expect  an  efficient and  accurate
composite prediction \eqref{equ_esp_QOI}.


%
\subsection{Supervised regression}\label{regressor} 
In  order to  enable XMA  predictions at  points $\etav_*$  outside the  training set,  a supervised
regression method is used to reconstruct the weighting functions $w_m(\etav_*)$.  More precisely, we
introduce supervised regressors for estimating the  cost functions $\cf_m$ at point $\etav_*$, which
are subsequently used to compute the weights.  \newline

In supervised  regression, a  set of  input-output pairs (the  training data  set) is  provided, the
assigned goal being  to learn the function that maps  inputs to outputs \cite{RusselNorvig_2020_AI}.
Many   well-known    model   class   algorithms    fall   into   this   category:    linear   models
\cite{Bishop_2006_PR&ML}, Support  Vector Machines (SVM) and  kernel methods \cite{Cortes_19995SVM},
Gaussian   Process   Regression   \cite{Williams_2006GPR},   Ensemble   Methods   based   on   trees
\cite{Breiman_1996Bagging} or Neural Networks \cite{Hinton_2006},  to cite just a few. \newline

In the rest of the study, we use Random Forests (RF) \cite{Breiman_2001RF}, which are well-suited to
large datasets  and highly non-linear  problems.  In the present  calculations we use  the RF
  package  available  in the  \emph{scikit-learn  python}  module.   The  algorithm relies  on  four
  hyperparameters:  (\emph{i}) the  number  of trees,  (\emph{ii}) the  maximum  number of  features
  evaluated  during a  splitting,  (\emph{iii})  the criterion  considered  for  node splitting  and
  (\emph{iv}) the minimum number of samples in a leaf.  The number of trees is fixed to a high value
  of 300,  while the other three  hyperparameters are optimized by  grid search. The grid  search is
  performed from a $K$-fold splitting with $K=10$.\newline

We train $\NM$ RF regressors modeling the relationship between the cost function $g_m$ and the features $\etav^{(m)}$ for each  of the component models.
Input-output                  pairs                   of                  the                  form:
$\C_m      =       \left\{      (\etav_{d}^{(m)},      g_m(\Mouts^m;\etav_{d}^{(m)},\datas_d,\hypar)
\right\}_{d=1}^{\ND}$, $m\in\{1,\cdots,\NM\}$ are used for the training.


For predictions, either on scenarios on which we  have reference data, or also on new scenarios, the
features   at   the   $\NP$   cell   nodes   to  be   predicted   are   collected   in   a   dataset
$\left\{  \etav_{j}^{(m)} \right\}_{j=1}^{\NP}$.   This  dataset can  be  used as  input  to the  RF
regressors to make predictions on $\cf_m$ and obtain the model weights.

As a result of the regression  process, RF approximations $\widetilde{g}_m(\etav^{(m)})$ of the cost
function are eventually obtained for each component model,
which are ultimately used to estimate the models' weights:
\begin{equation}\label{approx_weight_equ}
  \widetilde{w}_m(\etav^{(m)})          =         \frac{\widetilde{g}_m          (\etav^{(m)})}{\DPS
    \sum_{j=1}^{\NM}\widetilde{g}_j(\etav^{(j)})}, \quad m\in\{1,\cdots,\NM \} 
\end{equation}

In the following  numerical experiments, we observed that $g_m$  and $\widetilde{g}_m$ may sometimes
take values  approaching the  machine zero  for all $m$, making \eqref{approx_weight_equ}
becomes ill-conditioned, and possibly leading to incorrect results for the model weights.
%
Since low  values of  the cost functions  indicate overall  low confidence in  all of  the component
models, we introduce an  empirical lower bound to the cost functions $\coff$:  if the cost functions
for all models are below  $\coff$ at a given point of the feature space,  the XMA weights are simply
returned to the uniform choice $w_m=1/\NM$.  We conducted a sensitivity analysis on the value of the
cutoff-limit   $\coff$   and   we   observed   no  significant   influence   on   the   result   for
$\coff \in [0.001,0.15]$.  The value $\coff=0.001$ is thus retained for the rest of the study.

\subsection{Input features}\label{section_InputFeat}
Switching from  the space of geometrical  coordinates to a space  of input features is  an effective
method for generalizing model prediction to unseen geometries.   In this work, we select a subset of
10  features among  those initially  proposed by  Ling and  Templeton \cite{Ling_2015}.   The latter
define       a       feature       space       whose      elements       are       the       vectors
$\etav = (\expcomp_1,...,\expcomp_h,...,\expcomp_{10})^T$.
The  list of  input features  used in  this work  is reported  in Table  \ref{table_features}, where
$\vel_i$ denotes  the mean  velocity component  in the  $i$th space  direction, $\pres$  the average
pressure,  $\Ot$ the  mean rotation  rate, $\St$  the mean  strain rate,  $k$ the  turbulent kinetic
energy, $\varepsilon$ the turbulence dissipation rate, $\rho$ the fluid density, $\nu$ the kinematic
viscosity,  $\nu_T$ the  eddy viscosity,  and $||\cdot||$  is the  Frobenius norm.   Note that  some
turbulence models,  e.g. the Spalart-Allmaras  model considered later in  this work, do  not provide
estimates of the turbulent  kinetic energy $k$.  In such cases, the feature  is simply excluded from
considerations.

\begin{table}[]
\centering
\resizebox{1.1\textwidth}{!}{%
\begin{tabular}{|c|c|c|c|c|c|}
\hline
Feature & Description & Formula & Feature & Description & Formula \\ \hline
$\expcomp_1$ & Normalized $Q$ criterion & $\dfrac{||\Ot||^2 - ||\St||^2 }{||\Ot||^2 + ||\St||^2}$ & 
$\expcomp_6$ & Viscosity ratio & $ \dfrac{\nu_T}{100\nu + \nu_T}$ \\ \hline
$\expcomp_2$  &  Turbulence  intensity  &  $\dfrac{k}{  0.5\vel_i  \vel_i  +  k  }$  &  $\expcomp_7$
&   \begin{tabular}[c]{@{}l@{}}Ratio    of   pressure\\   normal   stresses    to\\   normal   shear
stresses\end{tabular} &  $\dfrac{\sqrt{\derp{\pres}{x_i} \derp{\pres}{x_i}}}{\sqrt{\derp{\pres}{x_j}
\derp{\pres}{x_j}} + 0.5\rho\derp{\vel_k^2 }{x_k}}$ \\ \hline
$\expcomp_3$    &    \begin{tabular}[c]{@{}l@{}}Turbulent     Reynolds\\    number\end{tabular}    &
$\min{\left(\dfrac{\sqrt{k}\lambda}{50       \nu},       2\right)}      $       &       $\expcomp_8$
&  \begin{tabular}[c]{@{}l@{}}Non-orthogonality  \\ marker  between  velocity  \\ and  its  gradient
\cite{Gorle_2012}  \end{tabular} &  $\dfrac{\left|\vel_k  \vel_l\derp{\vel_k}{x_l} \right|}{  \sqrt{
\vel_n   \vel_n  \vel_i\derp{\vel_i}{x_j}   \vel_m\derp{\vel_m}{x_j}  }   +  \left|   \vel_i  \vel_j
\derp{\vel_i}{x_j} \right|}$ \\ \hline
$\expcomp_4$  &  \begin{tabular}[c]{@{}l@{}}Pressure   gradient\\  along  streamline\end{tabular}  &
$\dfrac{ \vel_k\derp{\pres}{x_k} }{ \sqrt{\derp{\pres}{x_j}\derp{\pres}{x_j} \vel_i \vel_i} + \left|
\vel_l\derp{\pres}{x_l }\right|}$ & $\expcomp_{9}$ & \begin{tabular}[c]{@{}l@{}}Ratio of convection
to\\  production  of  $k$\end{tabular} &  $\dfrac{\vel_i  \derp{k}{x_i}}  {  |
\overline{u_j' u_l'}S_{jl} | + \vel_l \derp{k}{x_l} }$ \\ \hline
$\expcomp_5$  & \begin{tabular}[c]{@{}l@{}}Ratio  of turbulent\\  time scale  to mean\\  strain time
scale\end{tabular} & $\dfrac{ ||\St|| k}{ ||\St|| k + \varepsilon }$ & $\expcomp_{10}$ 
&   \begin{tabular}[c]{@{}l@{}}Ratio   of   total   Reynolds\\   stresses   to   normal\\   Reynolds
stresses\end{tabular} & $ \dfrac{||\overline{u_i' u_j'}||}{ k + ||\overline{u_i' u_j'}|| } $ \\
\hline

\end{tabular}%
}
\caption{ List  of input  features used  in this study.}\label{table_features}
\end{table}

\section{Simulation setup and reference data}\label{section_Refdata}

\subsection{Test case description}\label{case_setup}
\begin{figure}
\centering
    \includegraphics[width=0.5\linewidth]{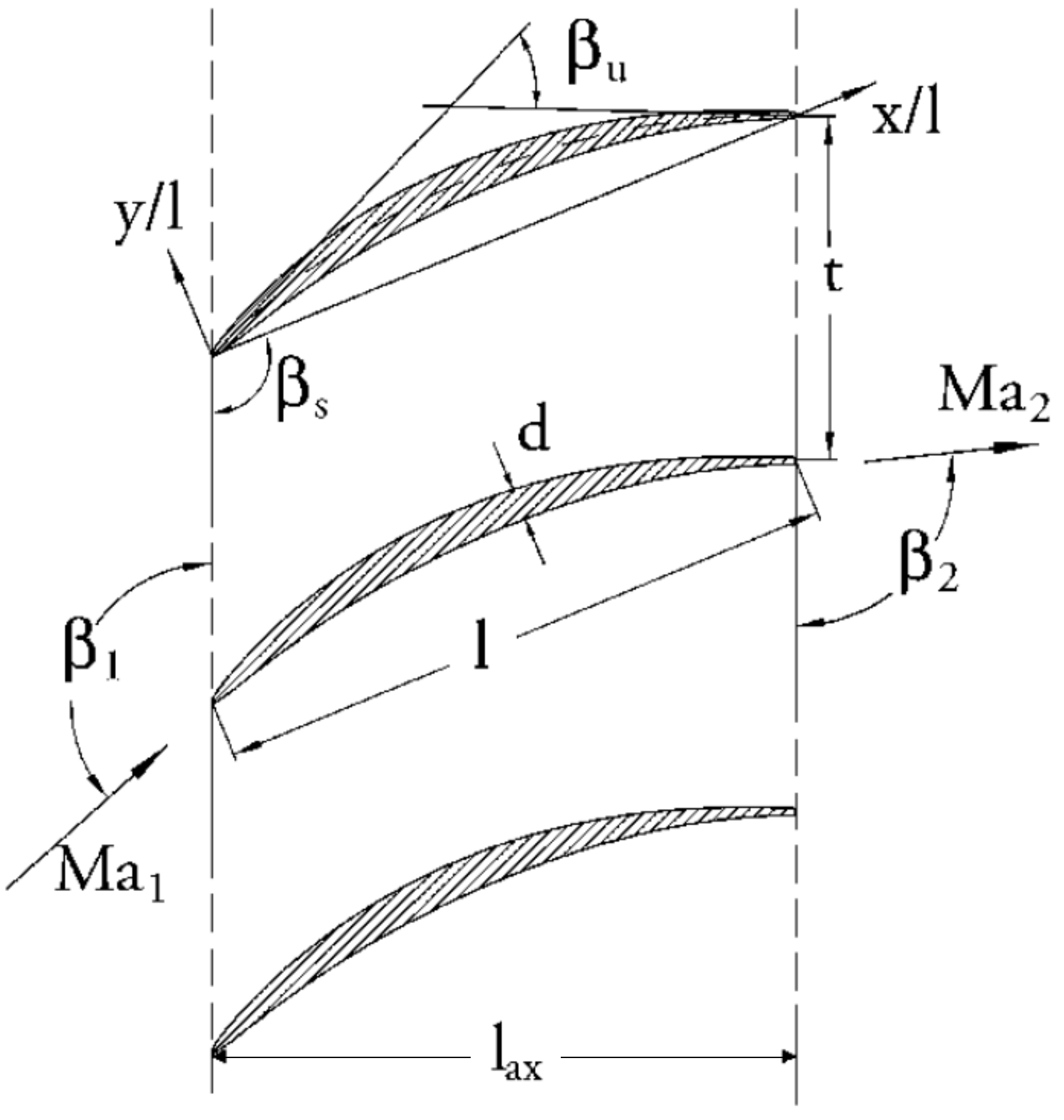}
 %
\caption{Sketch and design conditions of the NACA $65$ V$103-220$ linear compressor cascade. Sketch copied on from \cite{hilgenfeld_2003NACA65sansbarres}. }\label{fig_chapCaseSetup_presentationNACA65} 
\end{figure}

In the  following, the XMA  algorithm is demonstrated  for a flow  problem of practical  interest in
turbomachinery, the  turbulent flow through a  compressor cascade.  Specifically, we  model the flow
around  the  NACA $65$  V$103-220$  linear  compressor cascade,  widely  studied  in the  literature
\cite{leipold2000influence,        hilgenfeld_2003NACA65sansbarres,        iseler2006investigations,
  zaki2010direct,  leggett2016detailed} and  already  used as  a  test case  in  our previous  works
\cite{MZB_2020,MZB_2022}.  The cascade  is representative of the mid-span section  of a stator blade
in  a highly  loaded axial  compressor \cite{leipold2000influence}.   The blade  aspect ratio  being
$h/l = 1.36$, it  has been observed from the oil flow visualizations  performed on the blade surface
\cite{Bell_1995ASME} that the  flow "can be considered two-dimensional in  the mid-span section" for
the range  of Mach and Reynolds  numbers considered, which justifies  the present choice of  2D RANS
simulations.      The     cascade     geometry     and     nomenclature (taken     from     Ref.
\cite{hilgenfeld_2003NACA65sansbarres})               are                displayed                in
Fig. \ref{fig_chapCaseSetup_presentationNACA65}.

\begin{table}
\centering
\begin{tabular}{lcccc}
    \hline\hline
    Scenario            & $S_1$ & $S_2$ & $S_3$  & $S_4$ \\
    \hline\hline
    $\beta_1$ & 36.99\textdegree & 39.97\textdegree & 44.09\textdegree & 49.2\textdegree \\
    $Ma_1$ & 0.654 & 0.674 & 0.666 & 0.65 \\
    $Re_1$ & 302K & 302K & 298K & 289K \\
    Tu (\%) & 2.9 & 3.3 & 3.2 & 3.5  \\
    \hline\hline
  \end{tabular}
  \caption{Flow conditions for various compressor cascade scenarios.}\label{table_pres_scenarii}
\end{table}
\subsection{RANS solver and computational grid}\label{subsection_Setup}
The   simulations   are   carried  out   with   the   CFD   solver   $elsA$,  developed   by   ONERA
\cite{Cambier_2013_elsA}. The code solves the steady compressible RANS equations for Newtonian ideal
gases by means of  a cell-centered finite volume method on multi-block  structured grids. The upwind
scheme of Roe with  second-order MUSCL extrapolation is used for  approximating the inviscid fluxes,
and a  Gauss second-order scheme is  used for the viscous  fluxes.  The solution is  advanced to the
steady state by using the first-order backward Euler scheme and local time stepping.

The  computational domain  contains a  single blade  profile, and  periodic boundary  conditions are
applied at the upper  and lower boundaries to simulate an infinite cascade.  The domain extends from
$0.5$  axial chord  upstream of  the  leading  edge to  $1.0$  axial chord  downstream the  trailing
edge. The top  and bottom boundaries are separated  by a distance equal to  $0.59$ axial chord, which
also  represents the  gap  between neighboring  blades. No-slip  adiabatic  boundary conditions  are
applied at the blade wall, and characteristic conditions based on the Riemann invariants are imposed
at the inlet and  outlet boundaries. At the inlet, the total pressure,  enthalpy and angle of attack
are prescribed,  whereas a constant  static pressure is enforced  at the outlet.   The computational
grid is  composed by six matching  blocs, for a total  number of $30,880$ cells.  
The  near-wall grid resolution leads  to an average height  of the first
cell closest to the  wall (in wall coordinates) such that  $y^+ < 1.0 $ on both  the suction and the
pressure side of  the blade.  We verified that,  with such a resolution, the  numerical solution had
reached satisfactory mesh-independency.   Convergence to the steady state is  assumed when the $L_2$
norm of the residuals has been reduced by six orders of magnitude with respect to the initial value.

\subsection{Turbulence models}\label{subsection_RANSmodels}
The XMA predictions reported in Section \ref{section_Results} are constructed from a multi-model
ensemble of four concurrent linear-eddy-viscosity turbulence models (LEVM), selected amidst the most
widely used in industrial  practice.  Baseline RANS simulations of the  NACA $65$ V$103-220$ cascade
are conducted for each of the component LEVM  and for all scenarios in $\{S_1,S_2,S_3,S_4\}$, i.e. a
total  of 16  baseline RANS.   Such  simulations constitute  the  components of  the subsequent  XMA
procedure.   An additional  explicit algebraic  Reynolds stress  model (EARSM)  is used  to generate
reference data for training and validation,  as discussed in Section \ref{section_trainingsets}. The
reader is referred to the original articles cited in the following for more details about the models
in use.

\subsubsection{Component turbulence models}
\begin{description}

\item[{Spalart--Allmaras model \cite{spalart1994one} :}]  

  the most popular one-equation RANS model relies on a transport equation for an eddy-viscosity-like
  quantity $\tilde{\nu}$, which merges with turbulent viscosity $\nu_t$ far from the walls.

\item[{Wilcox $k-\omega$ \cite{wilcox2006turbulence}:}]

  the model  relies on transport  equations for  the turbulent kinetic  energy $k$ and  the specific
  dissipation rate $\omega$.  The eddy viscosity is then obtained as $\nu_t = {k} / {\hat{\omega}}$,
  with $\hat{\omega}$ a modified specific dissipation.

\item[{Launder--Sharma $k-\varepsilon$ \cite{launder1974KE}:}]

  the model  relies on transport equations  for the turbulent  kinetic energy $k$ and  the turbulent
  dissipation     rate     $\varepsilon$,    the     eddy     viscosity     being    obtained     as
  $\nu_t = {C_{\mu} k} / {\varepsilon}$, with $C_{\mu}$ a constant, generally taken equal to 0.09.

\item[{Smith $k-L$ \cite{smith_94KL}: }]

  the model was derived from the $k-kL$ model  of \cite{smith_90KKL} with the aim of simplifying the
  wall functions in the  $kl$ equation.  The model uses two transport equations  for $k$ and for the
  turbulence length scale $L$ . The turbulent  dissipation $\varepsilon$ is connected to $L$ through
  the following relation: $\varepsilon = (2k)^{3/2}/{B_1l}$.
\end{description}

\subsubsection{Reference model}  \label{section_refmodel} In addition  to the component  LEVM models
used in the XMA, we also consider a reference model relying on a different constitutive relation for
the  Reynolds stress  tensor, i.e.   displaying a  major structural  difference in  its mathematical
formulation compared  to LEVMs.   EARSM are expected  to provide a  more accurate  representation of
turbulence anisotropy  and rotation effects compared  to LEVM.  For  that reason, the model  is here
introduced for the  double purpose of generating training  data for the XMA and  assessing the model
prediction.   This allows  generating any  amount of  training data  for any  QoI and  for any  flow
scenario,  enabling  detailed parametric  analyses  and  solution  assessment (reported  in  Section
\ref{section_Results})  that  could be  hardly  achieved  with  the limited  high-fidelity  datasets
available in  the literature.  Precisely,  we consider  in the following  the EARSM $k-kL$  model of
\cite{bezard_2005_EARSMkkl}. The  latter models the  anisotropy tensor through  the Wallin-Johansson
formulation \cite{wallinjohansson_2000},  alongside Smith's $k-kL$ transport  equation for computing
the turbulent length and velocity scales \cite{smith_94KL}. \newline



\subsection{Training data sets}\label{section_trainingsets}
High-fidelity data are  collected for inferring the  XMA mixture weighting functions  in the feature
space. More precisely we consider four cascade operating conditions or scenarios, described in 
Table \ref{table_pres_scenarii}.
%
%
High-fidelity   DNS   and   LES   simulations   exist   in   the   literature   for   this   cascade
(e.g. \cite{zaki2010direct,leggett2016detailed}), but unfortunately  only limited results from those
datasets are publicly accessible.  This is why,  for the aims of the present proof-of-concept study,
we considered instead synthetic data sets generated through the reference EARSM $\kkl$ model.
The \emph{elsA} RANS solver  supplemented with the EARSM $k-kL$ model is  used to generate reference
full  fields  for various  flow  quantities  (including velocity,  static  and  total pressure,  and
temperature) for each scenario in $\{S_1,S_2,S_3,S_4\}$.  During preliminary tests (not reported for
brevity), the reference flow-field  quantities were used to investigate the  effect of inferring the
XMA model weights from various kinds of data.  Numerical tests showed that the total pressure is the
most informative  quantity.  Indeed, that total  pressure depends on both  dynamic and thermodynamic
quantities, and  therefore it  carries more  information about  the flow  than pressure  or velocity
separately,  and  particularly about  losses  in  the viscous  boundary  layers  and wake.   In  the
following, the total  pressure is retained as the  only observed quantity in the  data set $\datav$.
Reference fields of the remaining flow quantities are used for validation only.

\begin{figure}[H]
  \centering
  \begin{subfigure}[c]{\linewidth}
    \center
    \includegraphics[scale=0.35]{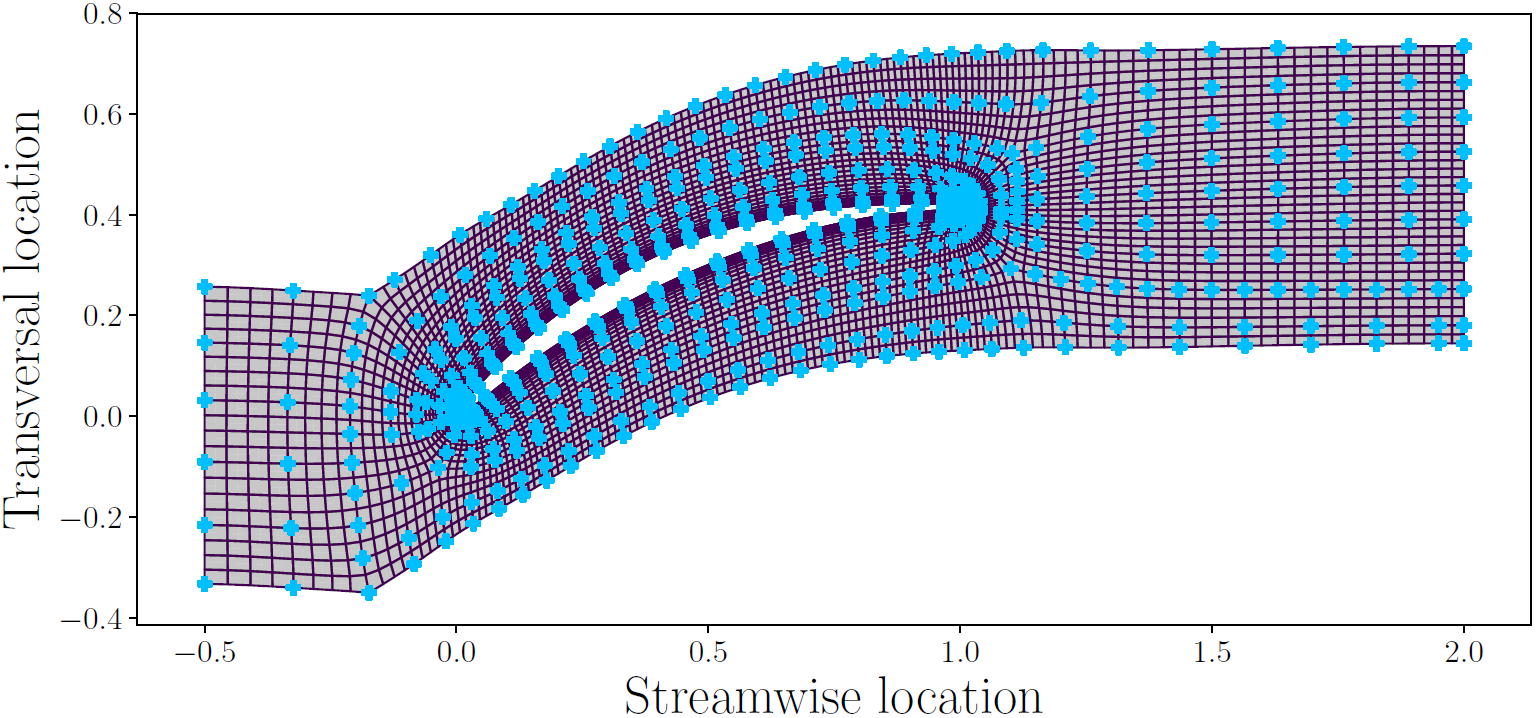}
  \end{subfigure}
  \captionsetup{justification=centering,margin=0cm}
  \caption{Locations of the  observation points for the small data regime. \\ For clarity, only one of four mesh vertices is represented.}\label{fig_subsetN=8}
\end{figure}

Two subsets of total  pressure data are then extracted from the  reference simulations, with the
aim  of investigating  the effect  of  training set  size on  the  learned XMA  weights.  Given  the
numerical  solutions available  at each  point of  the computational  mesh for  the four  scenarios a first data set is constructed by collecting total pressure data
at all mesh points, which we call the "big data" regime hereafter. The data are extracted at the mesh nodes, leading to a total of 40080 data.
We also consider a "small data" training regime corresponding to selecting one over 8 mesh points in each mesh direction, leading to a total of 820 data points.
  Figure \ref{fig_subsetN=8}  displays  the arrangement  of
numerical probes in the small data.  The data are  uniformly distributed across the grid,  and no attempt was  made to use optimal sensor
placement (OSP) techniques or prior physical knowledge on the flow, since it was beyond the scope of
the  present work.   Further  research on  optimal  sensor  placement is  warranted  in the  future.
\newline
%

In  the following,  numerical experiments  are  conducted by  training the  XMA weighting  functions
against observations for a single flow scenario  (i.e. by using total pressure
data for the chosen scenario), or for several scenarios simultaneously (\ie by using a concatenation of the total pressure datasets for the various scenarios).  
When multiple scenarios are
used to build the training set, the same observation points are considered for all scenarios, but other choices are possible.

\section{Results}\label{section_Results}
In this section the XMA is applied to predict flow past the NACA $65$ V$103$ cascade.

As a first  test, XMA is trained and  tested on the same flow scenario (namely  $\Scen_2$). A sensitivity
study of the  results to the hyperparameters of the  XMA model is carried out.  Afterwards,  XMA is used
for predicting a flow  scenario not used for training. More precisely,  we train XMA on three
scenarios  and  we  predict  on  the  fourth.   We  choose  to  train  the  algorithm  on  scenarios
$\{\Scen_2, \Scen_3, \Scen_4\}$  and predict on $\Scen_1$, in  order to have the angle  of attack of
the prediction  scenario outside the range  of the angles of  attack of the training  scenarios. As
described in Section \ref{case_setup}, $\Scen_1$ has a  severely off-design angle of attack and also
the lowest of the  four available scenarios, which makes this scenario an extrapolation configuration and a challenging  test case for assessing  XMA predictions
outside the training set.

\subsection{Training and prediction on scenario $\Scen_2$}
In this  section, the XMA  algorithm is trained  on EARSM $\kkl$  reference total pressure  data for
scenario $\Scen_2$  and applied  for the prediction  of the  full fields on  the same  scenario.  To
assess the  effect of the number of training data two  XMA models are constructed.   The first one,
noted $XMA_1$ is trained on the complete data set (\ie{} $40080$ data) and corresponds to the big data
regime; a  second model,  named $XMA_2$ uses  only $820$  data, which corresponds  to a  scarce data
regime.  Of note, the  second situation is the most likely to occur  in practice, especially in the
case of experimental measurements.

\begin{figure}
  \centering
  \begin{subfigure}[c]{0.48\linewidth}
    \center
    \includegraphics[width=\linewidth]{{Figures/chap_XBMA/NxN/L2_P2/Cartos/%
        L_2_P_2_dim_1_sigma_30_levelsPoids_CondCut_0.001_kepsls}.png} 
    \caption{ $\keps$ }\label{fig_NxN_carto_dim1_ke} 
  \end{subfigure}\hfill  
  \begin{subfigure}[c]{0.48\linewidth}
    \center
    \includegraphics[width=\linewidth]
  {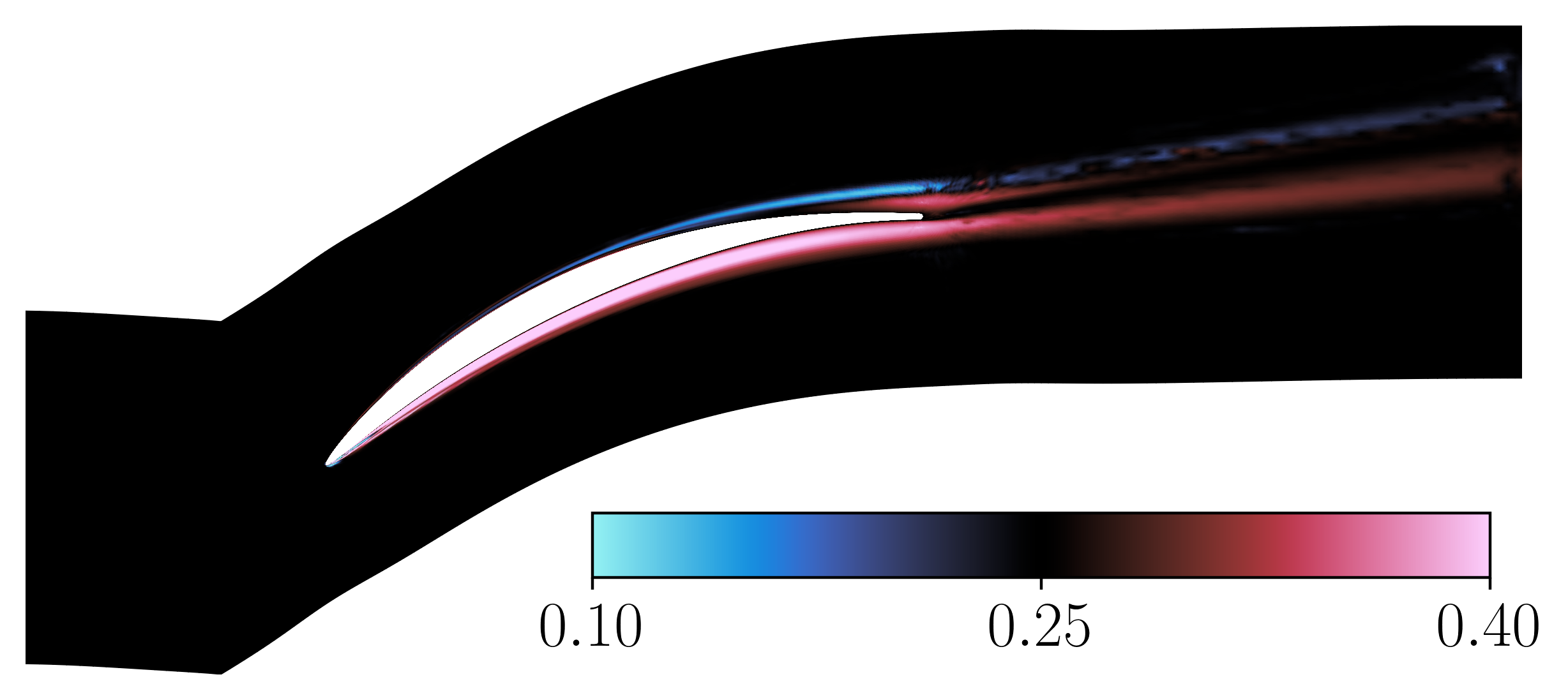}
    \caption{ Spalart-Allmaras}\label{fig_NxN_carto_dim1_sa} 
  \end{subfigure}
  \\[0.6cm]
  \begin{subfigure}[c]{0.48\linewidth}
    \center
    \includegraphics[width=\linewidth]
    {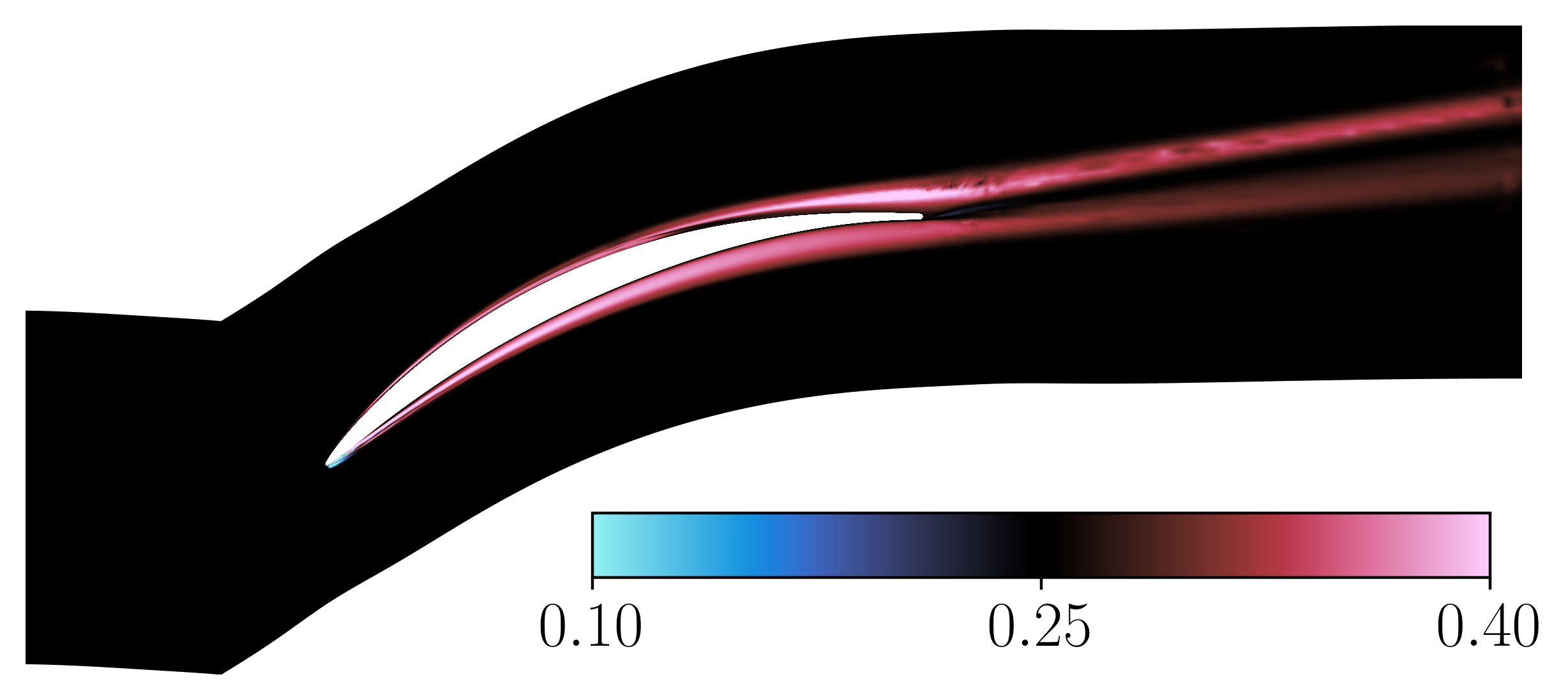}
    \caption{ $\kl$}\label{fig_NxN_carto_dim1_kl} 
  \end{subfigure}\hfill  
  \begin{subfigure}[c]{0.48\linewidth}
    \center
    \includegraphics[width=\linewidth]
    {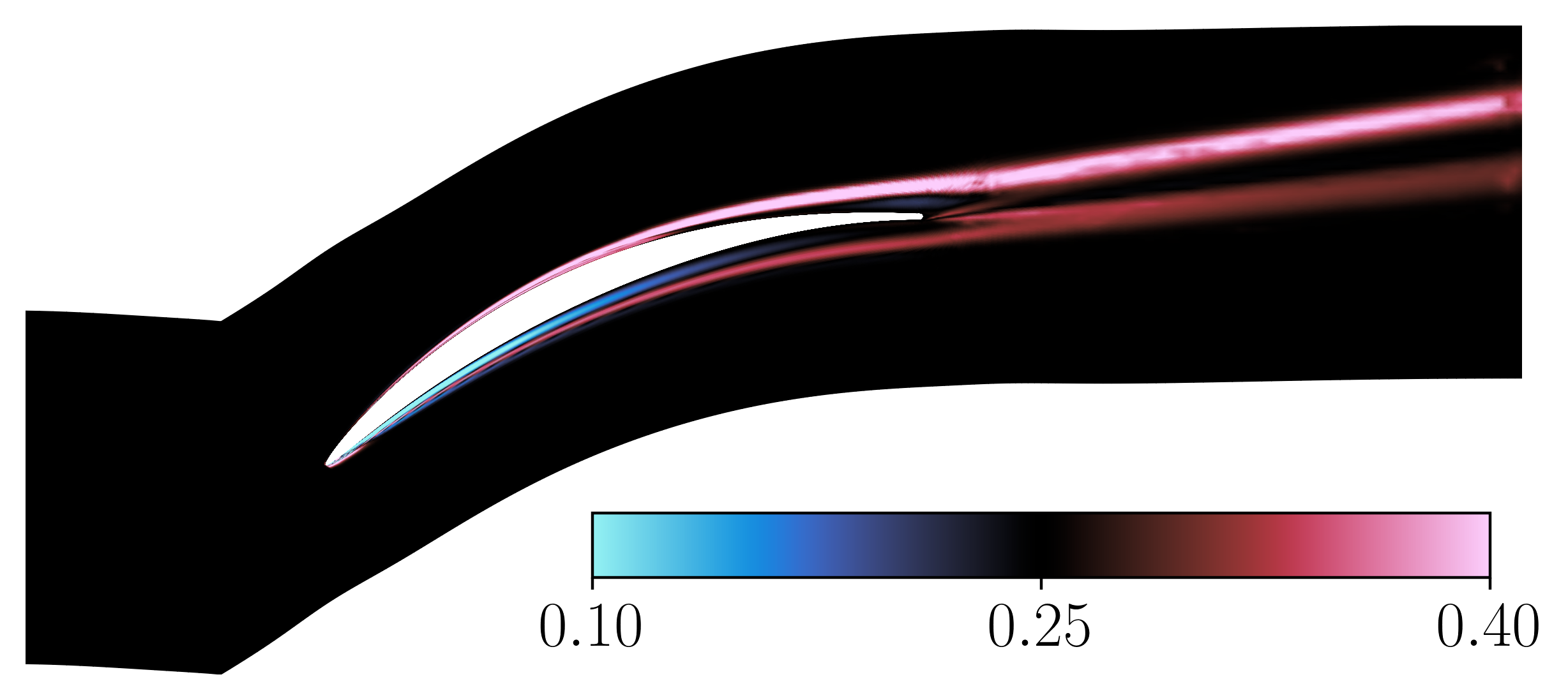}
    \caption{ $k-\omega$ }\label{fig_NxN_carto_dim1_ko} 
  \end{subfigure}
  \caption{Iso-contours of the $XMA_1$ weighting functions for the  four component RANS models. Training and prediction  on $\Scen_2$.}\label{fig_NxN_carto_dim1}
\end{figure}

Fig.   \ref{fig_NxN_carto_dim1} presents  the  iso-contours  of the  weighting  functions $w_m$  for
$XMA_1$ and the four component RANS models.  The predictions have been obtained in the feature space
and then brought back in the geometrical space $(x,y,z)$ to produce visualizations.  We observe that
all four models are assigned  weights equal to $1/4$ far from the blade,  i.e. in the potential flow
region.  This is consistent with the theoretical  expectation that RANS modeling does not affect the
potential region for the present flow with thin and attached boundary layers.  As a consequence, the
models are equally  likely to capture the reference  solution in that region.  On  the other hand,
the models weights  exhibit considerable differences in the  vicinity of the blade and  in the wake.
Fig.  \ref{fig_NxN_carto_dim1_ke}  shows that the $\keps$  model is generally associated  with lower
weights in  the wake and on  the suction and pressure  sides, meaning that $XMA_1$  has learned that
$\keps$ is the  less likely to accurately predict  the reference data in those  regions, compared to
the  other models.   The three  remaining  models are  globally  given higher  weights.  First,  the
Spalart-Allmaras  model (see  Fig.  \ref{fig_NxN_carto_dim1_sa})  is assigned  high  weights at  the
pressure side and, to a  minor extent, at the rear of the suction side,  close to the trailing edge,
while  the weight  tends  to go  down  to  $1/4$ in  the  wake.  Similarly,  the  $\kl$ model  (Fig.
\ref{fig_NxN_carto_dim1_kl}) is  given a  very high  weight at the  suction side  and a  rather high
weight  at the  pressure side.  Finally,  Fig.  \ref{fig_NxN_carto_dim1_ko}  presents the  weighting
function contours  for the $k-\omega$ model.   This model is given  the highest weight on  the first
half of the suction side,  which is then continued by a thin and detached band  of low weight on the
second half of the suction side.  A similar behavior is observed at the pressure side.

\begin{figure}
  \begin{subfigure}[c]{0.48\linewidth}
    \center
    \includegraphics[width=\linewidth]
    {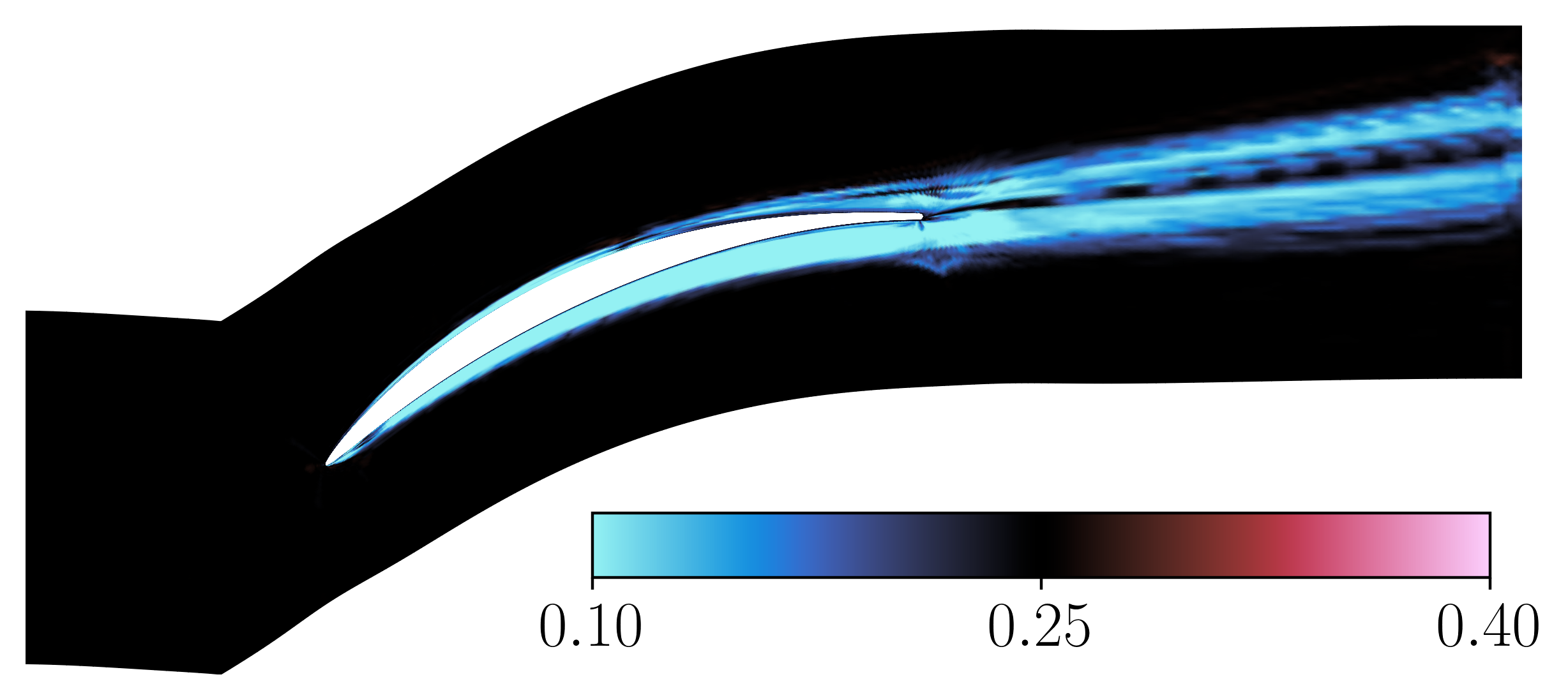}
    \caption{ $\keps$}\label{fig_NxN_carto_dim8_ke} 
  \end{subfigure}\hfill  
  \begin{subfigure}[c]{0.48\linewidth}
    \center
    \includegraphics[width=\linewidth]
    {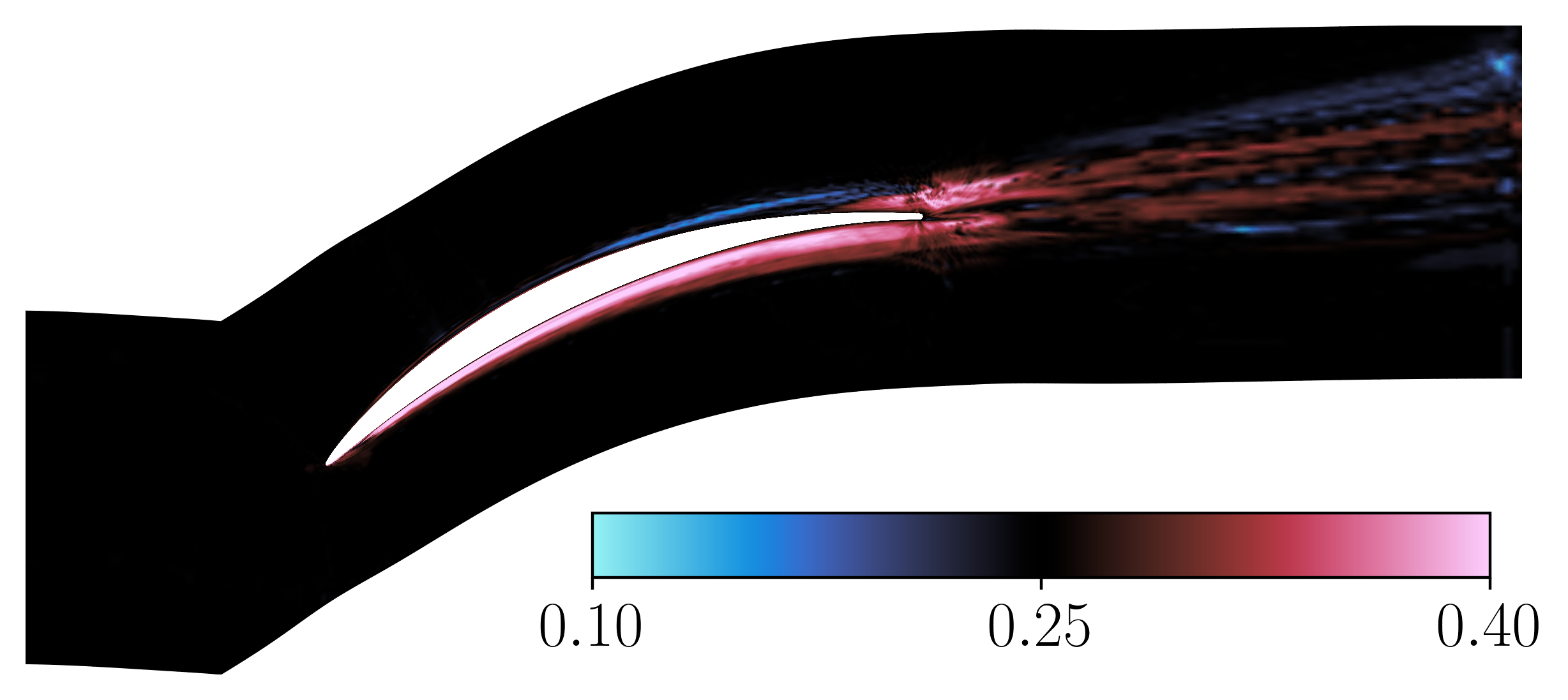}
    \caption{ Spalart-Allmaras}\label{fig_NxN_carto_dim8_sa} 
  \end{subfigure}
  \\[0.6cm]
  \begin{subfigure}[c]{0.48\linewidth}
    \center
    \includegraphics[width=\linewidth]
    {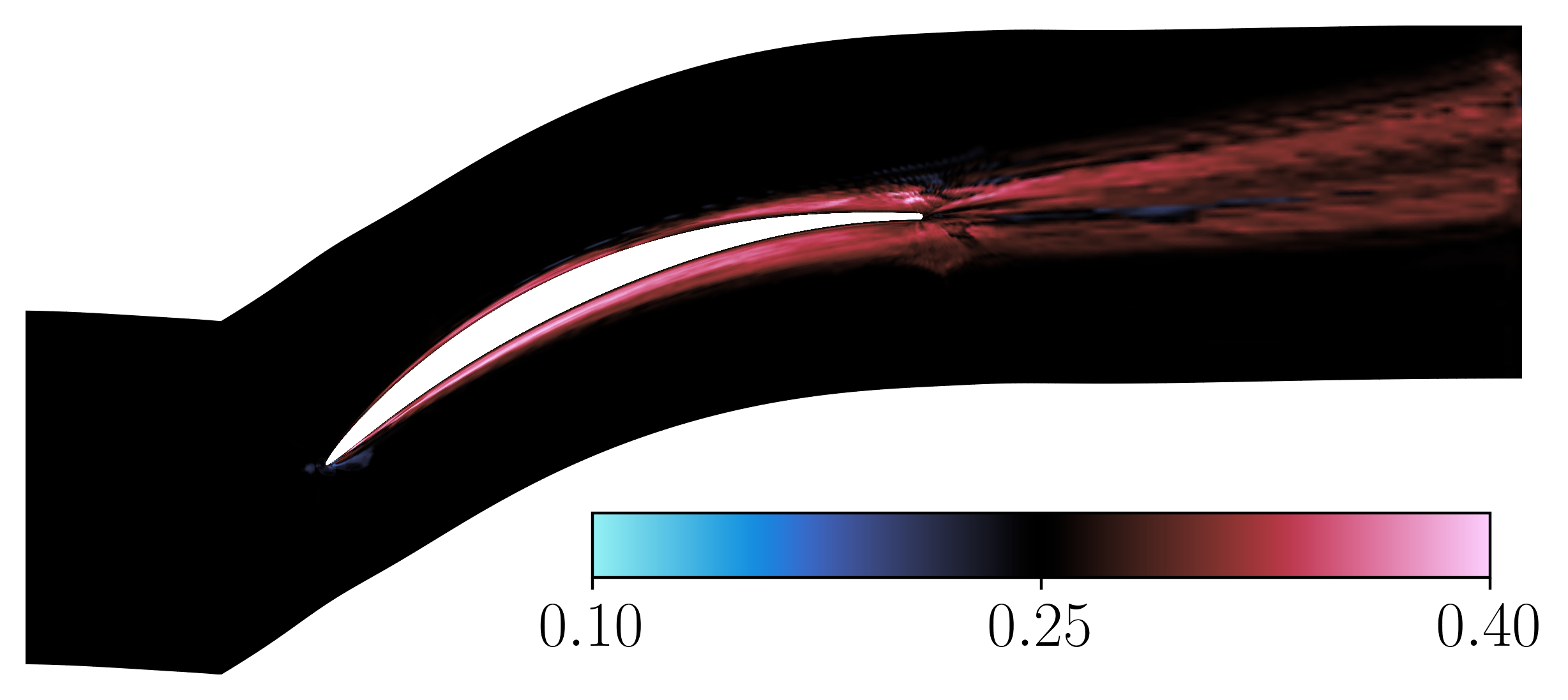}
    \caption{$\kl$}\label{fig_NxN_carto_dim8_kl} 
  \end{subfigure}\hfill  
  \begin{subfigure}[c]{0.48\linewidth}
    \center
    \includegraphics[width=\linewidth]
    {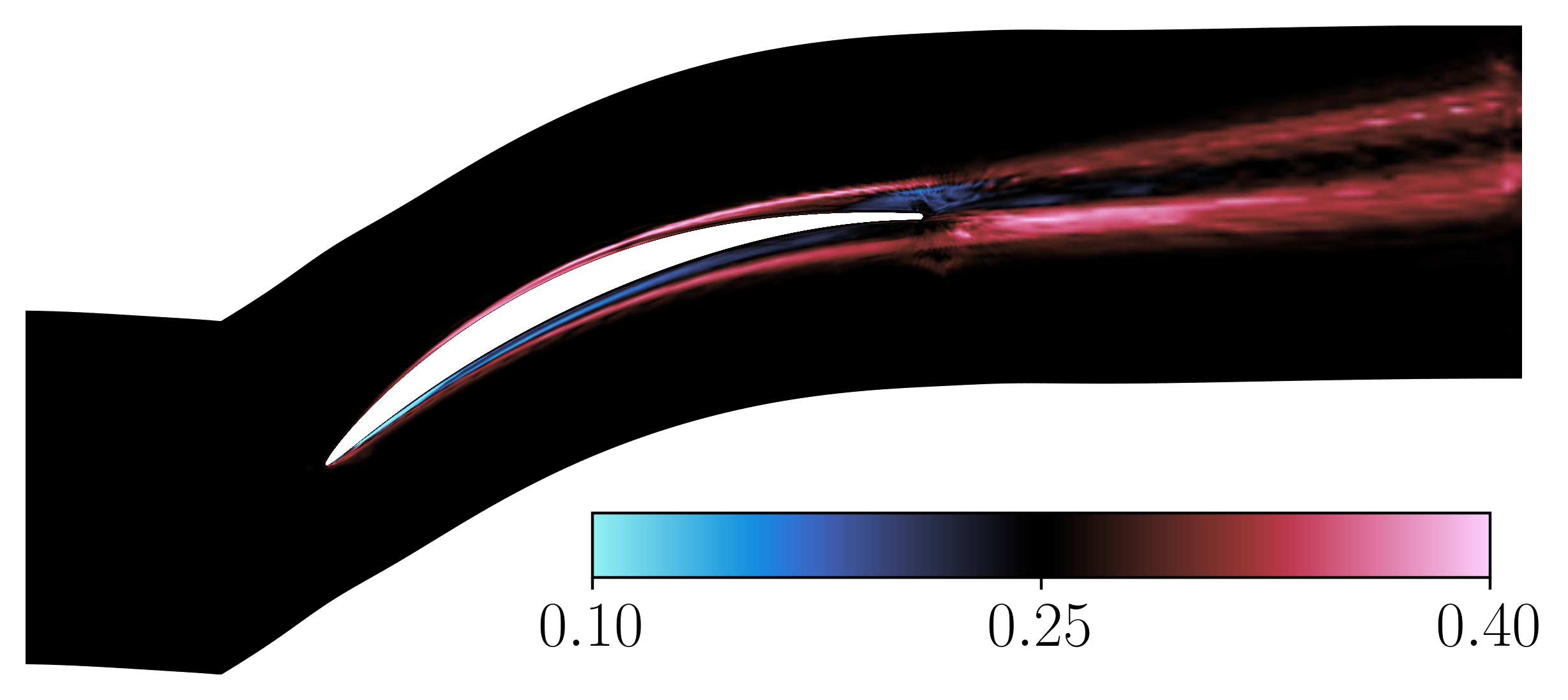}
    \caption{$k-\omega$}\label{fig_NxN_carto_dim8_ko} 
  \end{subfigure}
  \caption{ Iso-contours of the $XMA_2$ weighting functions for the  four component RANS  models. Training and prediction $\Scen_2$.}\label{fig_NxN_carto_dim8}
\end{figure}

Fig. \ref{fig_NxN_carto_dim8}  presents the  weighting function contours  for $XMA_2$,  trained with
only $820$  data.  Similarly  to Fig.   \ref{fig_NxN_carto_dim1_ke}, the $\keps$  model is  given an
overall lower weight than the other models. Overall, the contours are very similar to those obtained
in the big data regime ($XMA_1$), albeit a bit  noisier.  The observed noise is mostly caused by the
interpolation errors.
Nevertheless, the scarce data XMA results remain satisfactory and illustrate of potential of the proposed
methodology for real-world applications with limited training data.

\begin{figure} 
  \begin{subfigure}[c]{0.32\linewidth}
    \centering
    \includegraphics[width=\linewidth]
    {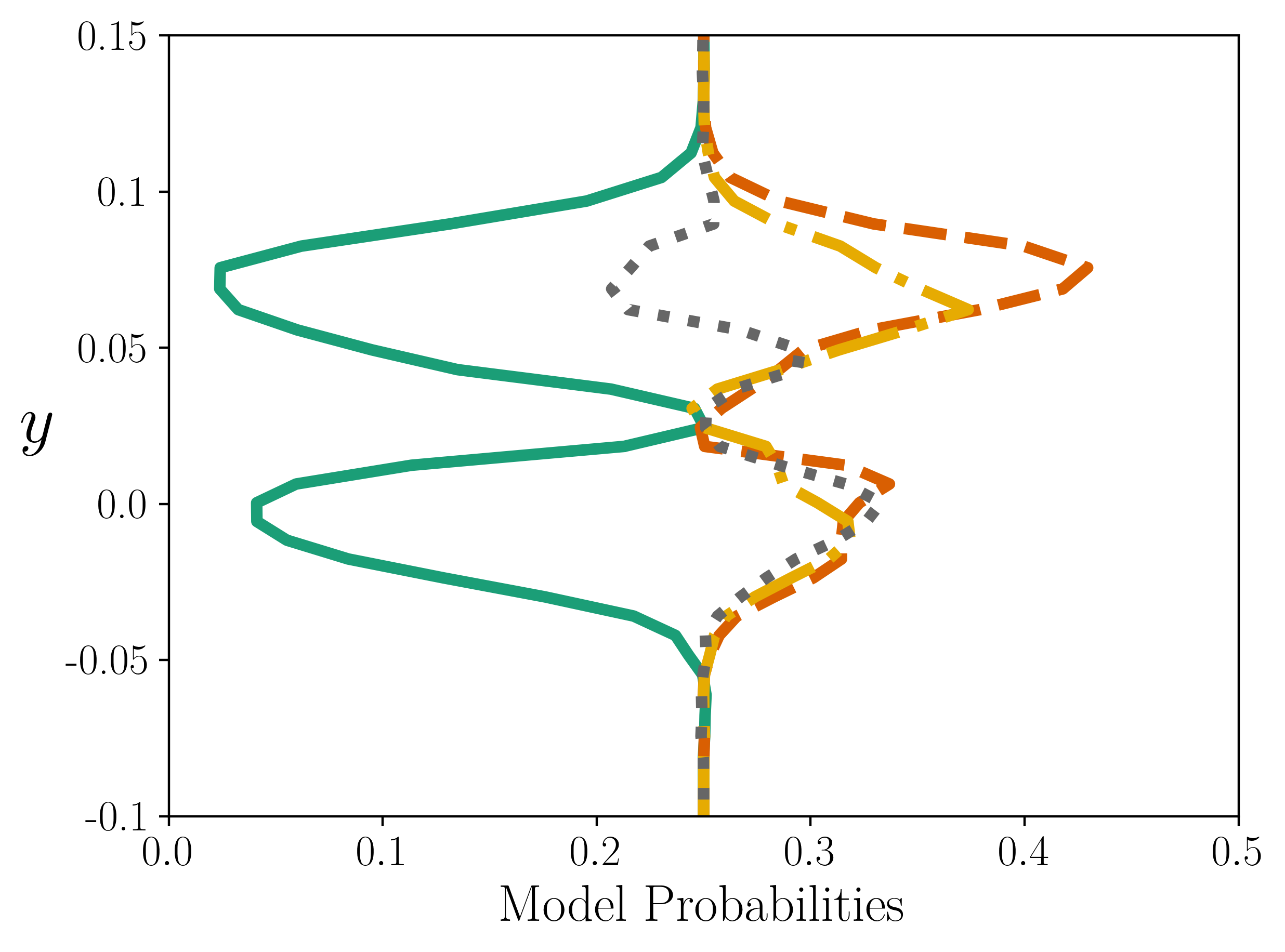}
      \captionsetup{justification=centering,margin=0cm}
    \caption{Model weights, $XMA_1$.}\label{fig_NxN_dim1_Sill1.2_poids} 
  \end{subfigure}
  \begin{subfigure}[c]{0.32\linewidth}
    \centering
    \includegraphics[width=\linewidth]
    {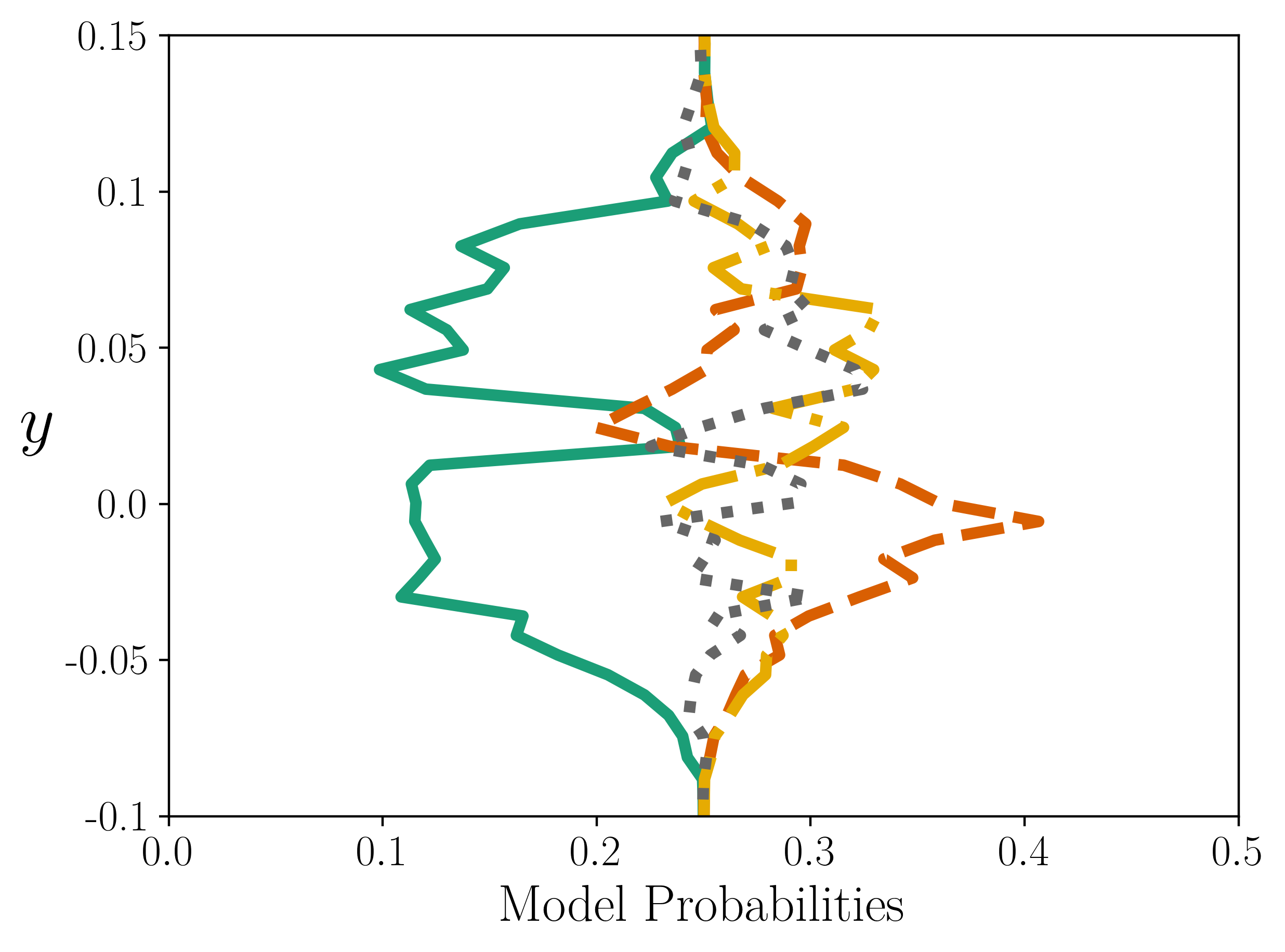}
    \captionsetup{justification=centering,margin=0cm}
    \caption{Model weights, $XMA_2$.}\label{fig_NxN_dim8_Sill1.2_poids} 
  \end{subfigure}
  \begin{subfigure}[c]{0.32\linewidth}
    \centering
    \includegraphics[width=\linewidth]
    {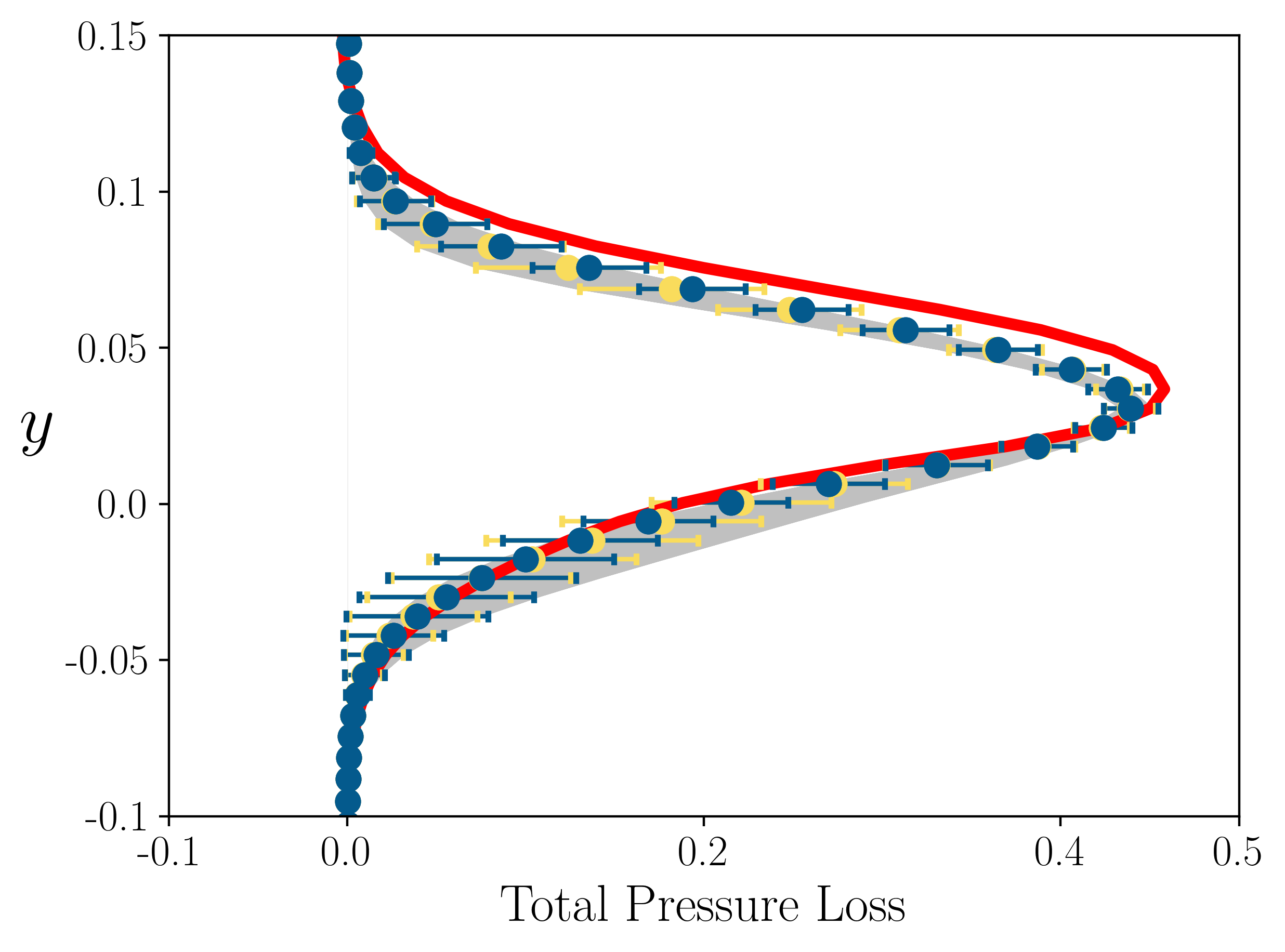} 
  \captionsetup{justification=centering,margin=0cm}
  \caption{Prediction}\label{fig_NxN_dim1_Sill1.2_pred} 
  \end{subfigure}
  \caption{Profiles of the weighting functions and of  the XMA prediction of the total pressure loss
    across the wake,  at streamwise location $\frac{x}{l}=1.20$. Solutions are  reported for the big
    data ($XMA_1$)  and scarce data  ($XMA_2$) regimes. Training  and prediction on  $\Scen_2$ total
    pressure data.  \\ 
    $\keps$   ($\protect\lsolid{green_Mod}$),   $k-\omega$   ($\protect\ldash{orange_Mod}$),   $\kl$
    ($\protect\ldashdot{yellow_Mod}$), Spalart-Allmaras ($\protect\ldott{grey_Mod}$), reference data
    ($\protect\lsolid{red}$), accessible  area \cbox{gray!50}, $\mean{\QOI}  \pm 2\sqrt{\var{\QOI}}$
    for    $XMA_1$   ($\protect\bp$),    $\mean{\QOI}    \pm    2\sqrt{\var{\QOI}}$   for    $XMA_2$
    ($\protect\yp$). }\label{fig_NxN_dim1_Sill1.2}
\end{figure}

We now evaluate more  quantitatively the quality of XMA training using different  amounts of data by
comparing the XMA prediction  of the quantity used for training (total  pressure) with the reference
data.  First, in Fig. \ref{fig_NxN_dim1_Sill1.2} we report profiles of the the model weights and the
corresponding   XMA   prediction   for   the    normalized   total   pressure   loss   (defined   as
$(Pt_{inlet}-Pt)/(Pt_{inlet}-P_{inlet})$ with $Pt$  the total pressure, $P$ the  static pressure and
$inlet$ denoting  the inflow boundary  of the computational domain)  at a given  streamwise location
across the wake ($\frac{x}{l}=1.20$).   Solutions are reported for both the big  data and the scarce
data    regimes.     Inspection of the weight profiles across the wake    (Figs
\ref{fig_NxN_dim1_Sill1.2_poids}  and  \ref{fig_NxN_dim8_Sill1.2_poids})  shows that  all  weighting
functions tend toward  the uniform weighting of $1/4$  in the potential region outside  the wake. As
observed previously in  the weighting functions contour  plots, the $\keps$ model is  assigned a low
weight throughout, whereas  the other RANS models are given  approximately equivalent higher values,
with the Spalart-Allmaras model being assigned somewhat lower  weight in the upper part of the wake,
consistently  with  the lower  weight  it  is assigned  in  the  suction-side boundary  layer.   The
scarce-data $XMA_2$ exhibits a qualitatively similar  behavior as $XMA_1$, with the different models
being assigned  weights of the same  order of magnitude, but  the weight profiles are  noisier. This
results from  the RF regressor  being less informed  in this case.   Despite that, both  $XMA_1$ and
$XMA_2$   provide   smooth   solutions,  rather   close   to   each   other,   as  shown   in   Fig.
\ref{fig_NxN_dim1_Sill1.2_pred}.   In  the  figure,  the two  XMA  predictions,  with  associated
variances  estimated from  Eq.  (\ref{equ_var_XBMA})  are compared  with  the  reference data.   The
grey-shaded region  represents the convex  hull of individual  predictions from each  component RANS
model. Since XMA constructs a convex linear combination  of the component models at each point, then
the XMA solution  must lie within that hull,  which is called hereafter the accessible  area. For the
QoI at stake the accessible area encompasses the  reference solution in the bottom part of the wake,
unlike the outer part of it.  The XMA solution captures well the reference in the bottom part despite
a large discrepancy  among the component models  (illustrated by the wide accessible  area) because it correctly assign high  local weights to the best performing models,
reducing the contribution of the worst performing one to the predictions.   In the upper part of the wake, all models
exhibit relative consensus on the wrong solution,  a known limitation inherent to mixture models. In
such a case,  the variances (a measure of model  consensus) are also small and do  not encompass the
reference either. However, XMA does a proper job of assigning higher weights to models providing the
best possible agreement with  the data.  The scarce-data $XMA_2$  tends to assign  more similar  weights to all  component models,
i.e.  it discriminates  less well  well-performing from  bad-performing models.   Nevertheless,
$XMA_2$ still provides improved performance overall, and much better performance than the worst model, showing that XMA prevents catastrophic loss
of accuracy with respect to the common-practice choice of a single (possibly wrong) RANS model.

Next, XMA is used  to reconstruct one-dimensional profiles of a flow quantity  not used for training
as a space-dependent linear combination of the baseline  RANS profiles for the same quantity and the
weighting functions.  We  focus more specifically on the prediction  of tangential velocity profiles
at chordwise location $x/l=0.9$, \ie{} at the rear of  the suction side, i.e. in a region of adverse
pressure gradient.
\begin{figure} 
  \begin{subfigure}[c]{0.32\linewidth}
    \centering
    \includegraphics[width=\linewidth]{{Figures/chap_XBMA/NxN/L2_P2/%
        Poids_Extrados_Vtang_-0.9_L2_P2_dim_1_sigma30_cut0.001}.png}
  \captionsetup{justification=centering,margin=0cm}
  \caption{Model weights, $XMA_1$.}\label{fig_NxN_dim1_Extra0.9_poids}
  \end{subfigure} \hfill
  \begin{subfigure}[c]{0.32\linewidth}
    \centering
    \includegraphics[width=\linewidth]
    {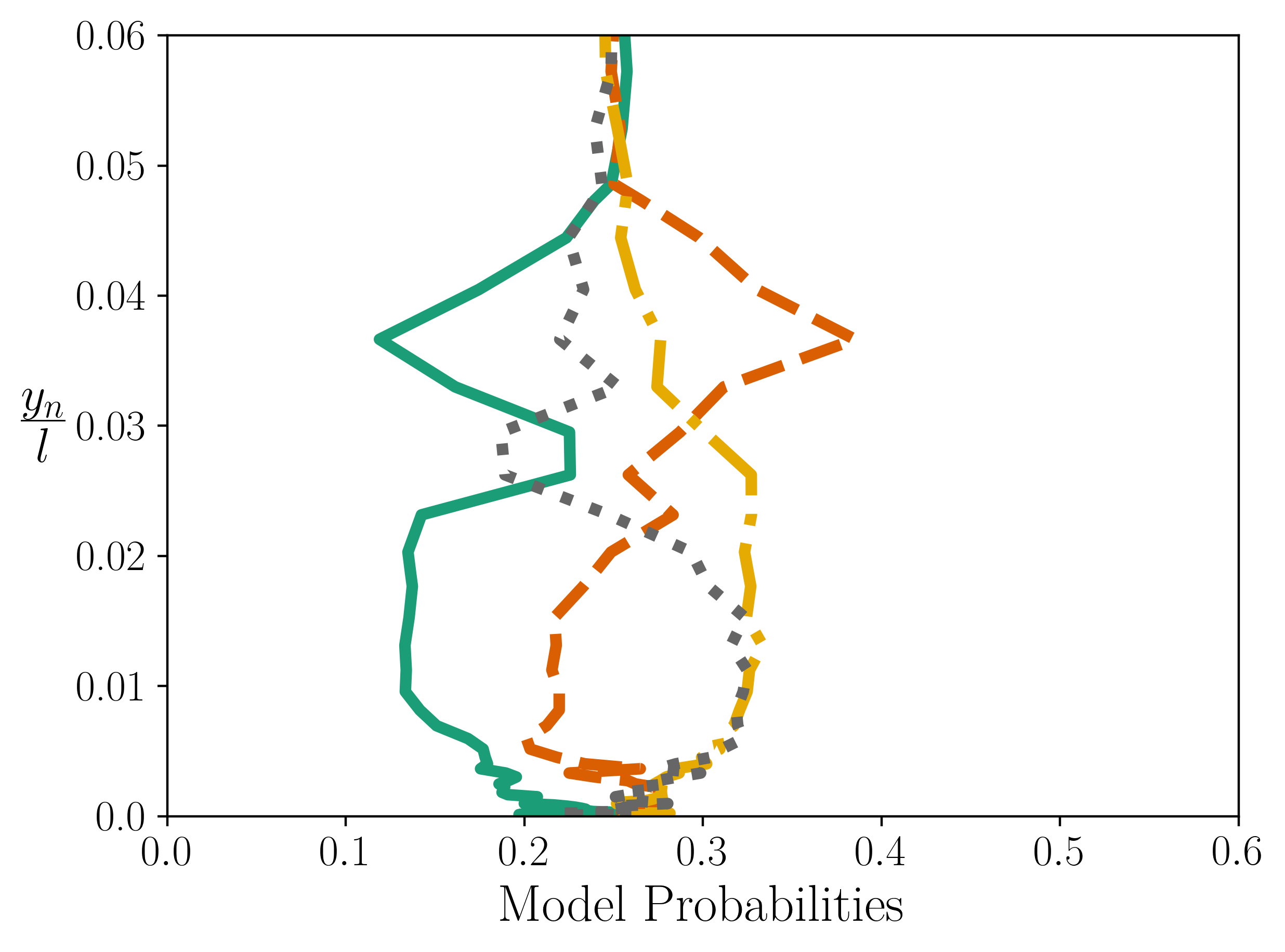}
  \captionsetup{justification=centering,margin=0cm}
  \caption{Model weights, $XMA_2$.}\label{fig_NxN_dim8_Extra0.9_poids}
  \end{subfigure}
  \begin{subfigure}[c]{0.32\linewidth}
    \centering
    \includegraphics[width=\linewidth]
    {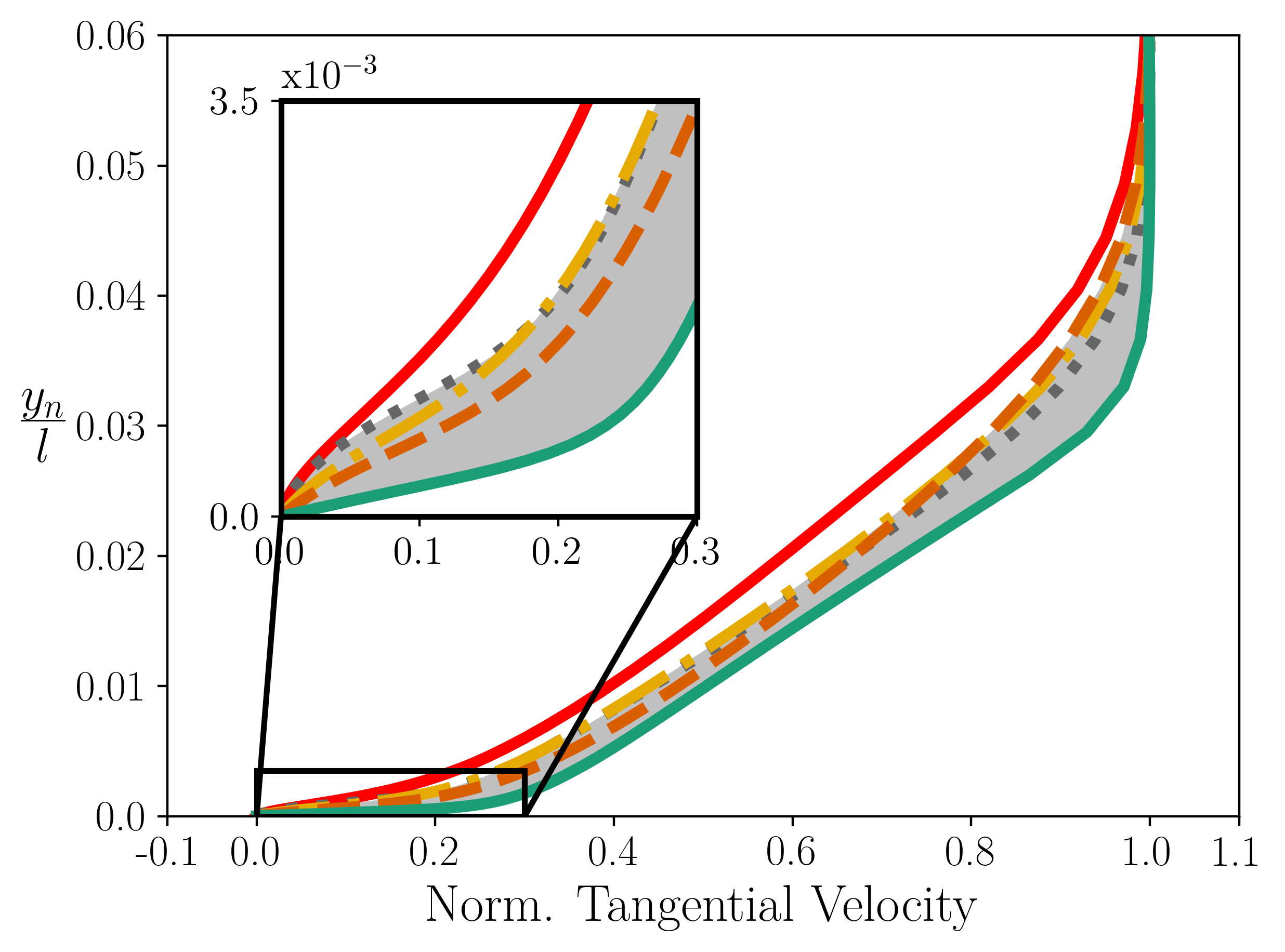}
  \captionsetup{justification=centering,margin=0cm}
    \caption{Prediction}\label{fig_NxN_dim1_Extra0.9_recons} 
  \end{subfigure}
  \caption{  Profiles of  the  weighting functions  (trained  on $\Scen_2$)  and  of the  tangential
    velocity  for  the   four  XMA  component  models,  at   streamwise  location  $\frac{x}{l}=0.9$
    ($\Scen_2$).  Solutions are reported for the big data ($XMA_1$) and scarce data ($XMA_2$)
    regimes.\\
    $\keps$   ($\protect\lsolid{green_Mod}$),   $k-\omega$   ($\protect\ldash{orange_Mod}$),   $\kl$
    ($\protect\ldashdot{yellow_Mod}$),  Spalart-Allmaras   ($\protect\ldott{grey_Mod}$),  accessible
    area            (             \cbox{gray!50}            ),             reference            data
    ($\protect\lsolid{red}$).}\label{fig_NxN_poids_Extra0.9.}
\end{figure}
Figure \ref{fig_NxN_poids_Extra0.9.}  shows the model  weight distributions for $XMA_1$  and $XMA_2$
obtained after  training on total  pressure data.  As  observed previously, the  weighting functions
have similar trends (noisier  for $XMA_2$) but they exhibit sharper  difference in the well-informed
$XMA_1$ than in $XMA_2$, which in turn tends to assign weights closer to the uniform distribution of
$1/4$.  Once again the  $\keps$ model is assigned much lower weights in  both cases, while $\kl$ and
Spalart-Allmaras are preferred in  the inner part of the boundary layer and  $k-\omega$ in the outer
part.  Figure \ref{fig_NxN_dim1_Extra0.9_recons} shows the  accessible area, the reference data, and
the individual  component RANS solutions for  the tangential velocity profiles  (normalized with the
velocity at boundary layer  edge $U_e$).  The XMA algorithm is expected to rank  the models in each region
according  to their  agreement  with the  data,  \ie{} model  weights  should be  a  measure of  the
predictive accuracy  of the corresponding  component model.   However, the weighting  functions were
learned from total  pressure data, so we may  wander if they still provide a  reasonable estimate of
model accuracy for  velocity profiles.  We observe  that $\keps$, the lowest weighted  model, is the
worst performing one in this case. The three other  RANS models are closer to the reference, but the
enlarged plot in  the inset shows that  the $k-\omega$ model is less  accurate than Spalart-Allmaras
and $\kl$ models close to the blade, and  it is assigned a lower weight accordingly.  Similarly, the
Spalart-Allmaras is less  accurate in the outer part  of the boundary layer, and  it is consistently
assigned a very  low weight in that region,  both in $XMA_1$ and $XMA_2$.  This  shows that, despite
XMA is not  directly trained on the predicted  QoI, still it provides reasonable  estimates of local
model performance.
\begin{figure}[H]
  \begin{subfigure}[c]{0.48\linewidth}
    \centering
    \includegraphics[width=\linewidth]
    {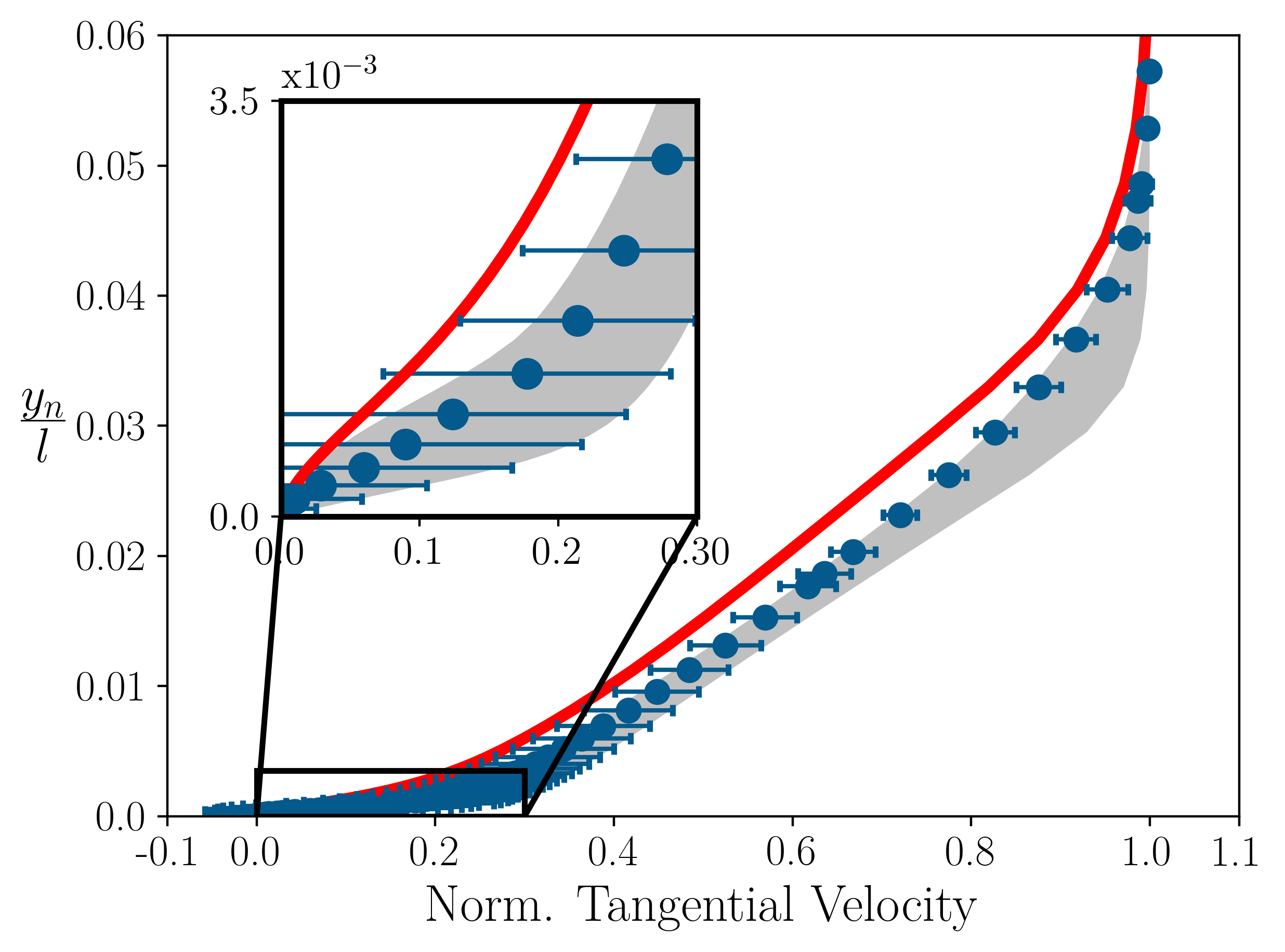} 
    \caption{$XMA_1$}\label{fig_NxN_dim1_Extra0.9_pred}
  \end{subfigure}
  \begin{subfigure}[c]{0.48\linewidth}
    \centering
    \includegraphics[width=\linewidth]
    {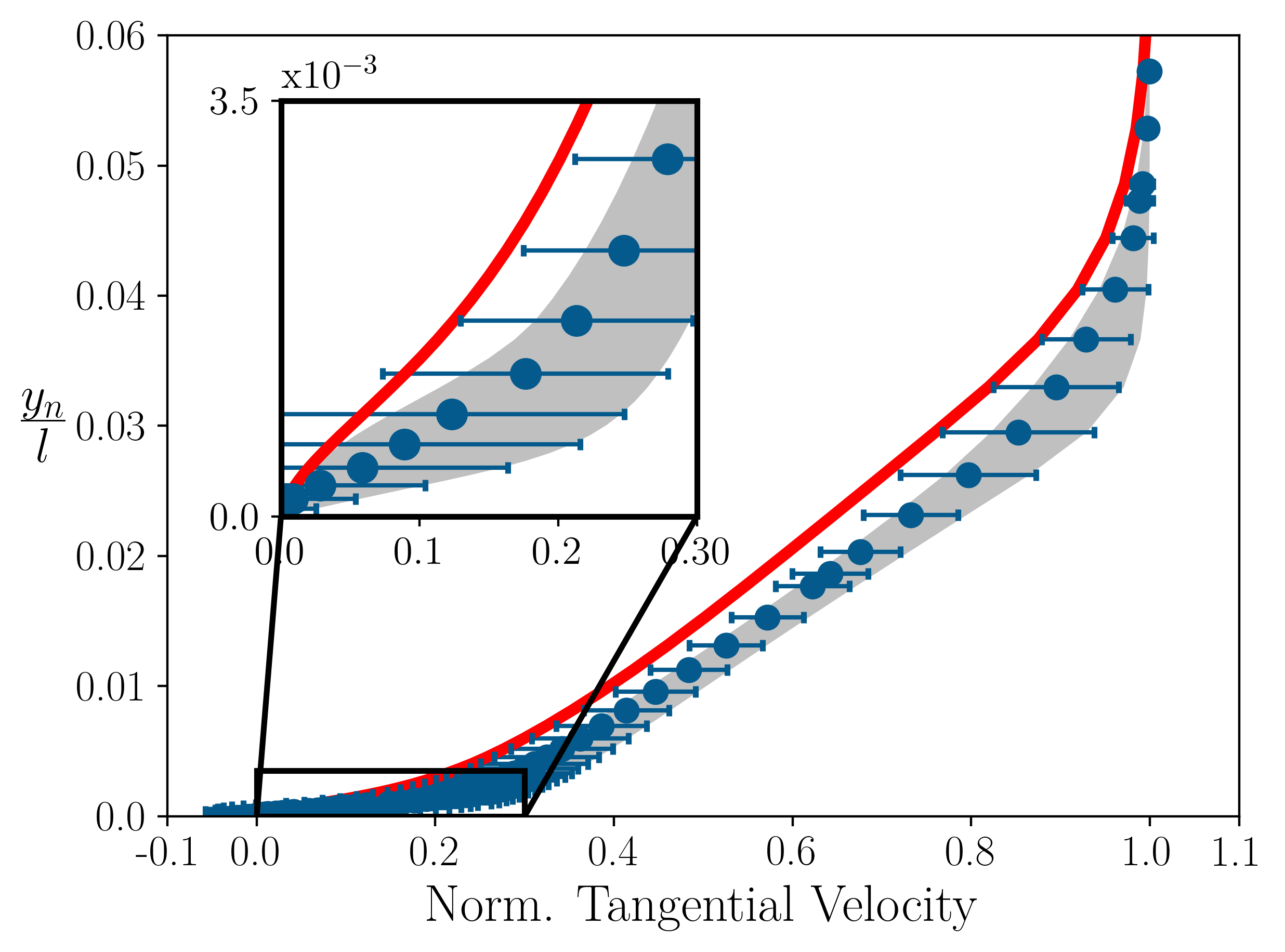} 
    \caption{$XMA_2$}\label{fig_NxN_dim8_Extra0.9_pred} 
  \end{subfigure}
  \caption{XMA prediction of the tangential velocity profile at $\frac{x}{l}=0.90$, $\Scen_2$,
    for $XMA_1$ and $XMA_2$.
    Reference  data  ($\protect\lsolid{red}$),  accessible  area  \cbox{gray!50},  $\mean{\QOI}  \pm
    2\sqrt{\var{\QOI}}$ ($\protect\bp$).}\label{fig_NxN_pred_Extra0.9} 
\end{figure}

In Fig.   \ref{fig_NxN_pred_Extra0.9} we present  the reconstructed tangential velocity  profiles at
the same streamwise  station, with error bars  corresponding to twice the variance.   The two panels
correspond      to     $XMA_1$      (Fig.     \ref{fig_NxN_dim1_Extra0.9_pred})      and     $XMA_2$
(Fig.  \ref{fig_NxN_dim8_Extra0.9_pred}),  respectively.  For  both  the  big  and the  scarce  data
regimes,  XMA predictions  are in  better agreement  with the  reference data  than individual  RANS
models, showing that the algorithm is correctly  preferring the best-performing model in each region
of the  boundary layer.   Further, the  variances provide an  estimate of  local model  consensus, a
measure of  the risk  of obtaining  a significantly  wrong solution if  a single  model was  used to
predict the flow.  In the present case, the  error bars encompass the reference data but, as already
observed for the total pressure loss profiles, the area accessible by the component models does not.
In other terms, the error bars must not be interpreted as the region were the true solution possibly
lies, but simply as a measure of the uncertainty in the choice of a best-performing model.  Of note,
XMA assigns weights in such  a way that  the mixture prediction tends  to the accessible  area limit
closer to the reference data.  This effect is stronger  for $XMA_1$ than $XMA_2$, as  previously observed for
the  total pressure  loss,  since bringing  more  information to  the training  set  makes XMA  more
selective  locally.  This  also  results in  larger  variances  for $XMA_2$,  since  all models  are
contributing  rather significantly  to the  mixture everywhere,  while in  $XMA_1$ only  one or  two
best-performing models are assigned high weights in each region.
%

\begin{figure} 
  \centering
  \begin{subfigure}[c]{0.50\linewidth}
    \centering
    \includegraphics[width=\linewidth,trim=0 0 0 0, clip]
    {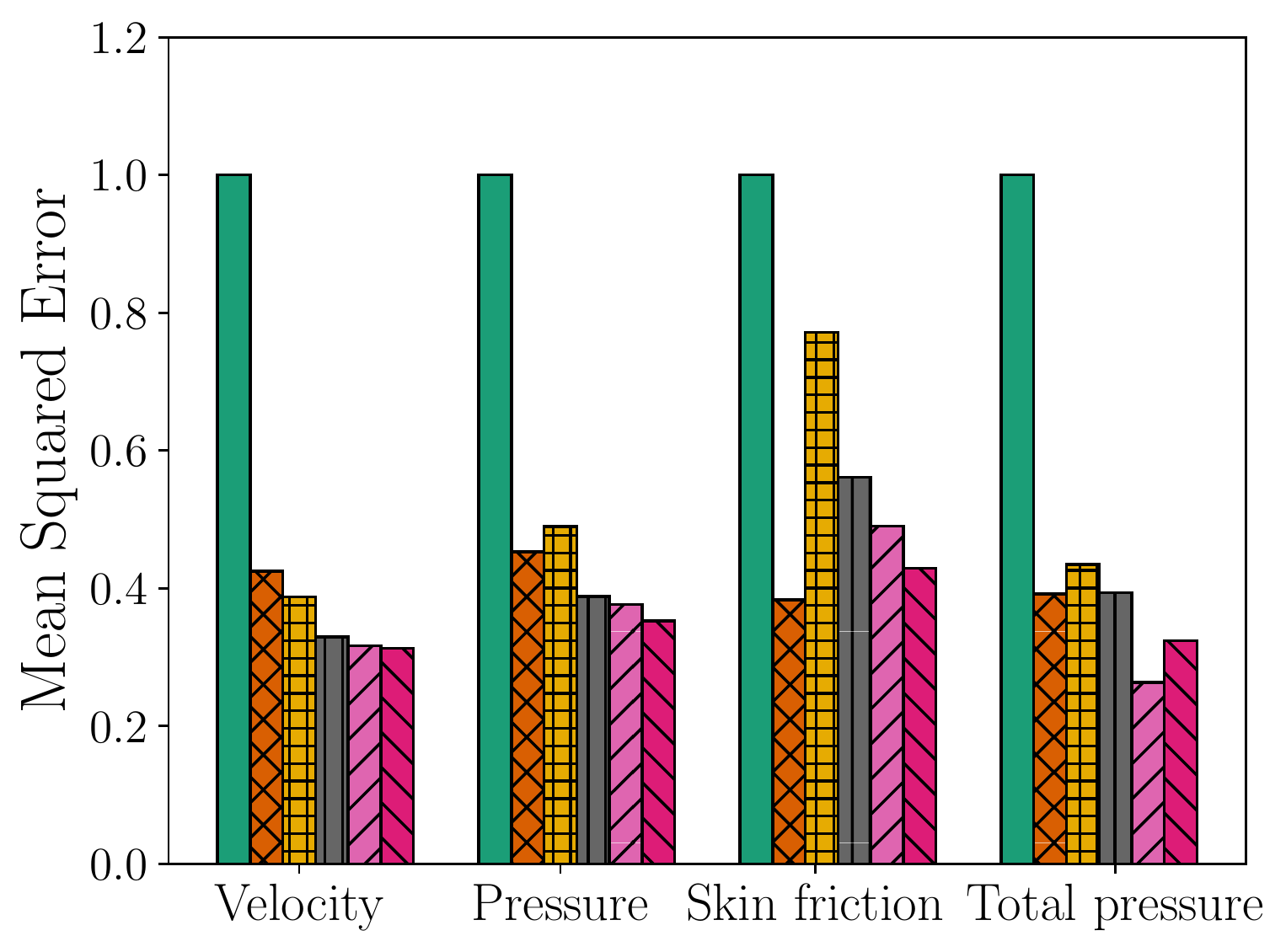}
  \end{subfigure}
  \caption{ Mean-squared errors for four QoI (normalized by the MSE of the $\keps$ model),
    $\Scen_2$. XMA is trained on $\Scen_2$. \\ 
    $\keps$  model  (\bprectKE),  $k-\omega$  model  (\bprectKO ),  $\kl$  model  (  \bprectKL)  and
    Spalart-Allmaras   model   (\bprectSA),   $XMA_1$  (\bprectXBMAA)   ,   $XMA_2$   (\bprectXBMAB)
  }\label{fig_NxN_MSE_adimKE} 
\end{figure}

To provide an overview  of the accuracy of XMA for various QoIs,  in Fig. \ref{fig_NxN_MSE_adimKE} we
report the global  mean squared errors (MSE) with  respect to the reference data  for four different
QoI (pressure, velocity, skin  friction and total pressure).  The quantities  are estimated at each
point of the full computational mesh using XMA, and the predicted values are compared with the reference
solution, except for  the skin friction that is  reconstructed only for mesh points  along the blade
wall.  Of  the four  quantities, three  were not  used for model  training. The  errors of  the four
baselines RANS models  are also reported for  comparison.  We observe that the  baseline RANS models
exhibit very different performance, even for a  relatively simple $2$-D configuration as the present
compressor  cascade.  Additionally,  model  accuracy strongly  depends  on the  QoI  at stake.   For
instance, the  Spalart-Allmaras model  provides closer  agreement with  the reference  for velocity and pressure,
while $k-\omega$  provides a  better estimate  of the skin  friction.  The  $\keps$ is
consistently less accurate than all other models for  all QoI, which is in accordance with the fact
that in  \Cref{fig_NxN_carto_dim1_ke,fig_NxN_carto_dim8_ke} this model is  systematically assigned a
lower  weight.  The  $\kl$ model  is  the second  least accurate  model  for the  skin friction,  in
contradiction  with \Cref{fig_NxN_carto_dim1_kl,fig_NxN_carto_dim8_kl}  where  the  model is  highly
likely  all over  the blade.  However, the  weights have  not been  trained on  skin friction  data.
Furthermore, the  $\kl$ model has been  found to be  particularly inaccurate close to  leading edge,
which radically  deteriorates its average  performance, although  the prediction is  reasonably good
elsewhere.  Turning now to XMA performance, we  see from Fig.  \ref{fig_NxN_MSE_adimKE} that the XMA
provides the most accurate  predictions for three of the considered quantities and  not only for the
one used for training, even if the error for  total pressure is lower than for the other quantities,
as expected.  For the skin friction, XMA performs  worst than $k-\omega$, but is still more accurate
than all the  other component models. The results  could be improved in the future  by improving the
selection and placement of training data.  A particularly encouraging result is that both $XMA_1$ and
$XMA_2$ are improving  the results, despite the  large difference in training set  size, opening the
way to the training of XMA from relatively scarce data sets (e.g. experimental datasets).

\begin{figure}
  \centering
  \begin{subfigure}[c]{0.48\linewidth}
    \center
    \includegraphics[width=\linewidth]
    {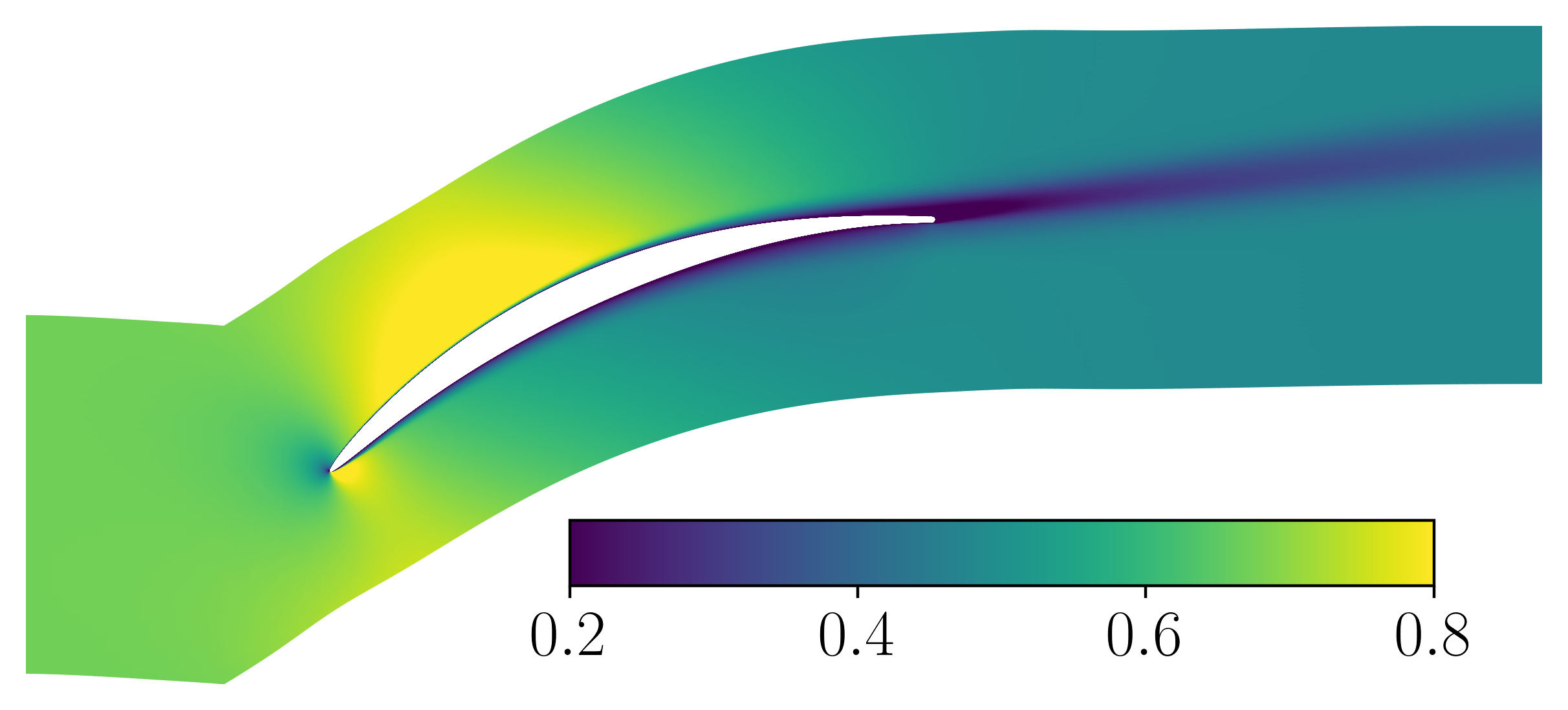}
    \caption{ Mean prediction.}\label{fig_NxN_dim8_2DMach_Esp} 
  \end{subfigure}\hfill  
  \begin{subfigure}[c]{0.48\linewidth}
    \center
    \includegraphics[width=\linewidth]
    {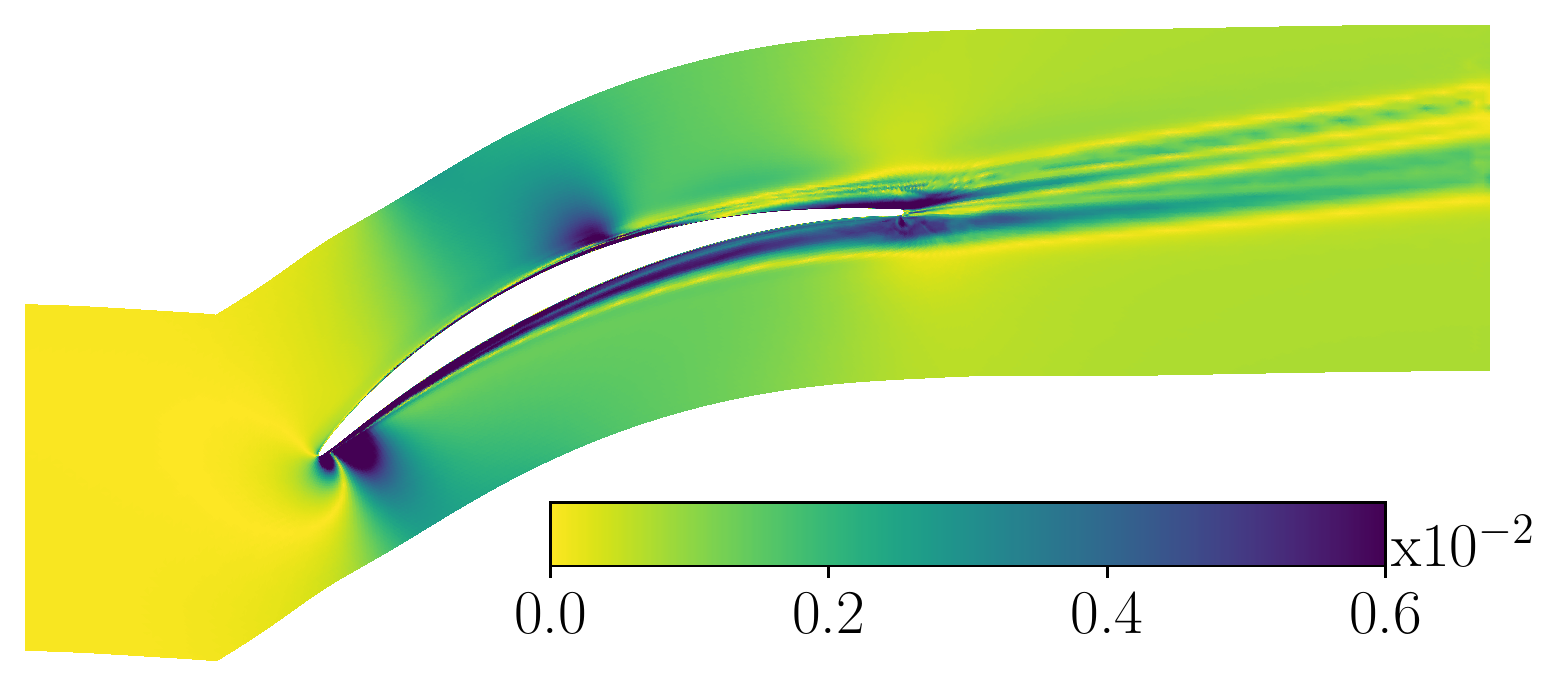}
    \caption{Prediction variance.}\label{fig_NxN_dim8_2DMach_Var}
  \end{subfigure}
  \caption{Iso-contours    of    the    Mach    number    field    predicted    with
    $XMA_2$ (training and prediction on $\Scen_2$).}\label{fig_NxN_dim8_2DMach}
\end{figure}

Finally, we  complete the  analysis of  this case by  presenting in  Fig.  \ref{fig_NxN_dim8_2DMach}
$2$-D  contour plots  of  the  expectancy and  variance  of the  Mach  number  field, obtained  with
$XMA_2$. Despite  the scarce data  used for  training, the $2$-D  fields are remarkably  smooth.  We
verified that the same consideration holds for  other QoI.  As expected, high variances are observed
in the near-wall  region and in the  turbulent wake, but also  in regions of the  external flow
more  strongly coupled  with  the viscous  layers,  typically, the  regions  with stronger  pressure
gradient directly affected by the boundary layer development.


\subsection{Prediction of an unseen scenario}\label{XBMA_subsection_PredS1_EARSM}
\begin{figure}
  \begin{subfigure}[c]{0.48\linewidth}
    \center
    \includegraphics[width=\linewidth]{{Figures/chap_XBMA/NxN/L234_P1/Cartos/%
        L_234_P_1_dim_1_sigma_30_levelsPoids_CondCut_0.001_kepsls}.pdf}
    \caption{ $\keps$ }\label{fig_NxN_carto_dim1_ke_L234_P1}
  \end{subfigure} \hfill
  \begin{subfigure}[c]{0.48\linewidth}
    \center
    \includegraphics[width=\linewidth]
    {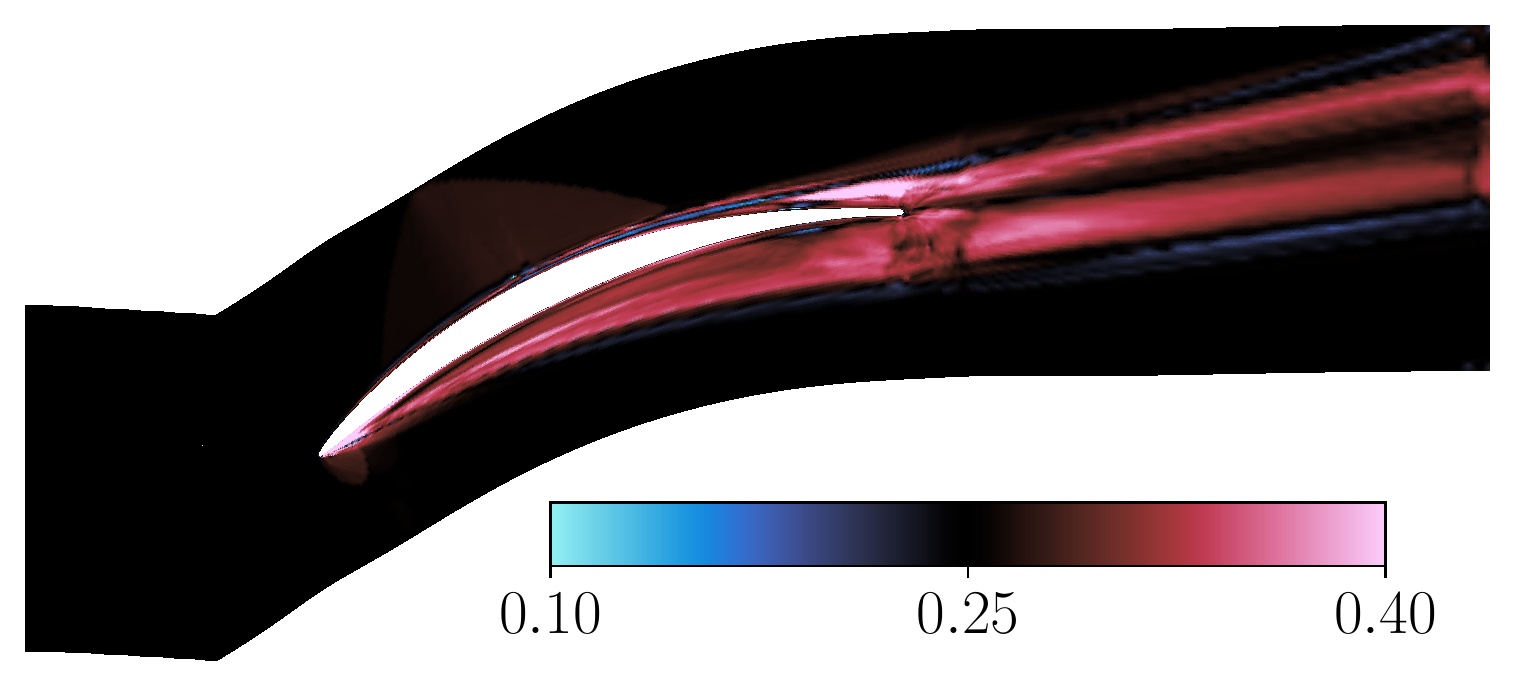}
    \caption{ Spalart-Allmaras }\label{fig_NxN_carto_dim1_sa_L234_P1} 
  \end{subfigure}
  \\[0.6cm]
  \begin{subfigure}[c]{0.48\linewidth}
    \center
    \includegraphics[width=\linewidth]
    {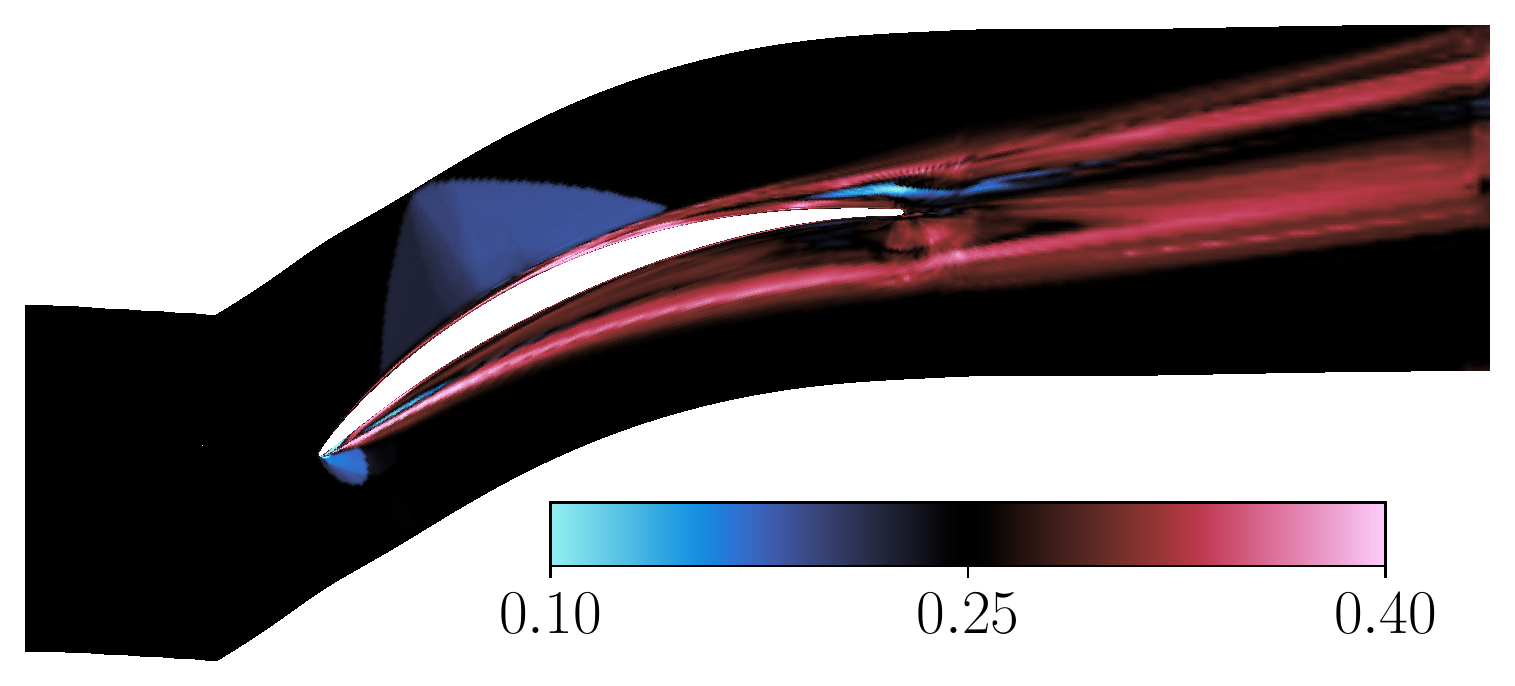}
    \caption{$\kl$}\label{fig_NxN_carto_dim1_kl_L234_P1} 
  \end{subfigure}\hfill
  \begin{subfigure}[c]{0.48\linewidth}
    \center
    \includegraphics[width=\linewidth]
    {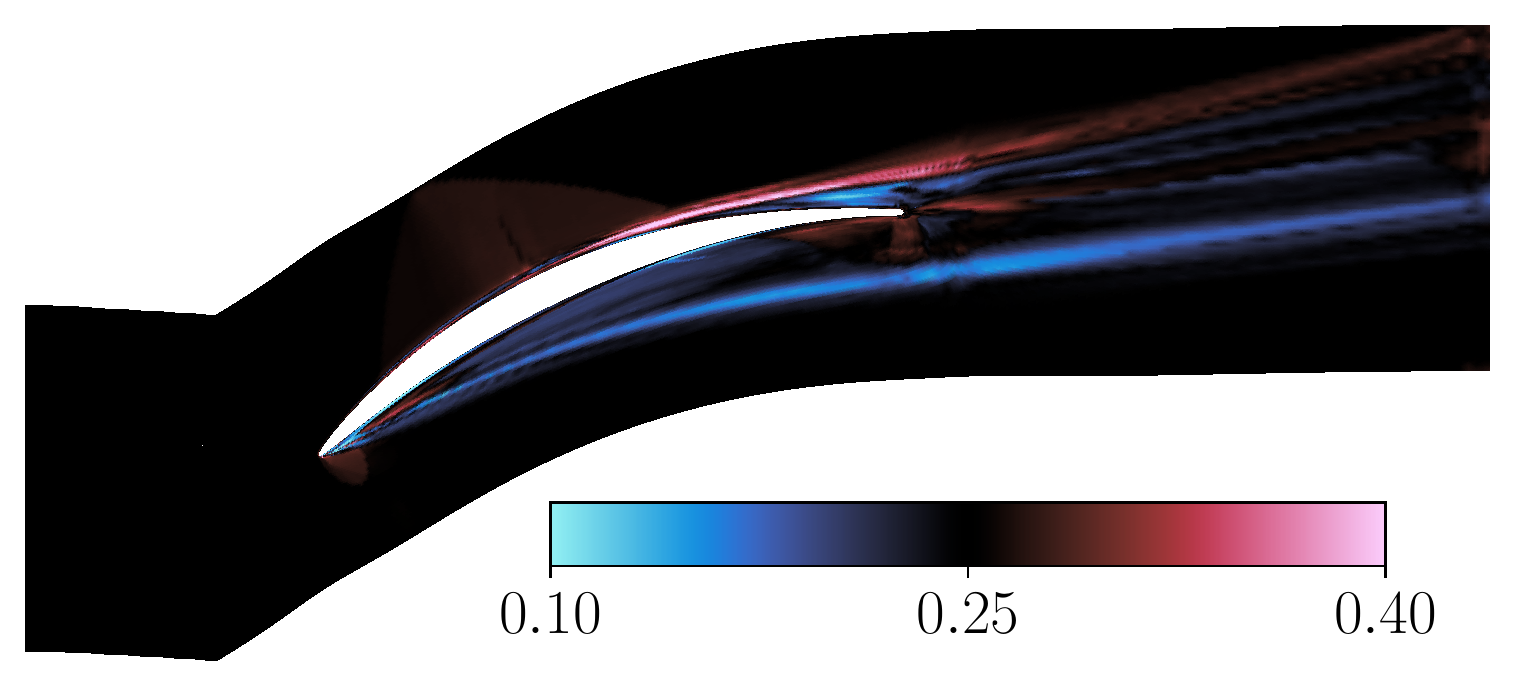}
    \caption{$k-\omega$ }\label{fig_NxN_carto_dim1_ko_L234_P1} 
  \end{subfigure}
  \caption{Isocontours of $XMA_1$  weighting functions for the four component  RANS models. Training
    on $\{\Scen_2,\Scen_3,\Scen_4\}$ and prediction on $\Scen_1$.}\label{fig_NxN_carto_dim1_L234_P1} 
\end{figure}

In  this Section  we evaluate  the  ability of  XMA at  predicting  an unseen  flow scenario.   More
precisely,   the XMA  algorithm   is  trained   simultaneously  with   data  extracted   from  scenarios
$\{\Scen_2,  \Scen_3,  \Scen_4\}$  and used  to  predict  $\Scen_1$.   The  training data  sets  are
constructed by concatenating total pressure data for the three scenarios.  We note again $XMA_1$ the
big  data regime,  with a  training set  corresponding to  full total  pressure data  for each  flow
scenario (\ie{}  $120240$ data in  total) and $XMA_2$  the scarce data  regime, with a  training set
constituted of $820$ data for each scenario (\ie{} $2460$ data in total).
%
Fig.  \ref{fig_NxN_carto_dim1_L234_P1} presents contour plots  of the four model weighting functions
$w_m$ in the  case of the $XMA_1$.  The maps  obtained for $XMA_2$ are very similar  and are omitted
for the  sake of brevity.   
As in
the preceding tests, $\keps$ model is assigned low  weights in all regions of interest, i.e.  in the
vicinity    of    the   blade    and    in    the    wake.     The   Spalart-Allmaras    model    in
Fig. \ref{fig_NxN_carto_dim1_sa_L234_P1} is assigned high weights on both pressure and suction side,
as well as in the  wake. The region close to the trailing edge  is particularly associated with high
weights, by  contrast with  the other  models that  are systematically  assigned lower  weights.  In
contrast with the prediction on  $\Scen_2$, Fig.  \ref{fig_NxN_carto_dim1_kl_L234_P1} shows that the
$\kl$  model is  assigned high  weight only  in the  outer parts  of the  suction and  pressure side
boundary layers, but an intermediate weight of $1/4$ on the pressure side and even low weight on the
suction side close  to the trailing edge.   Finally, $k-\omega$ model is assigned  higher weights at
the suction side, in the outer part of the boundary layer, and lower weights close to the wall inner
part.  At the pressure  side on the contrary, the $k-\omega$ model seems  to be better performing in
the inner part of the boundary layer rather than in the outer part.

\begin{figure}
  \begin{subfigure}[c]{0.32\linewidth}
    \centering
    \includegraphics[width=\linewidth]
    {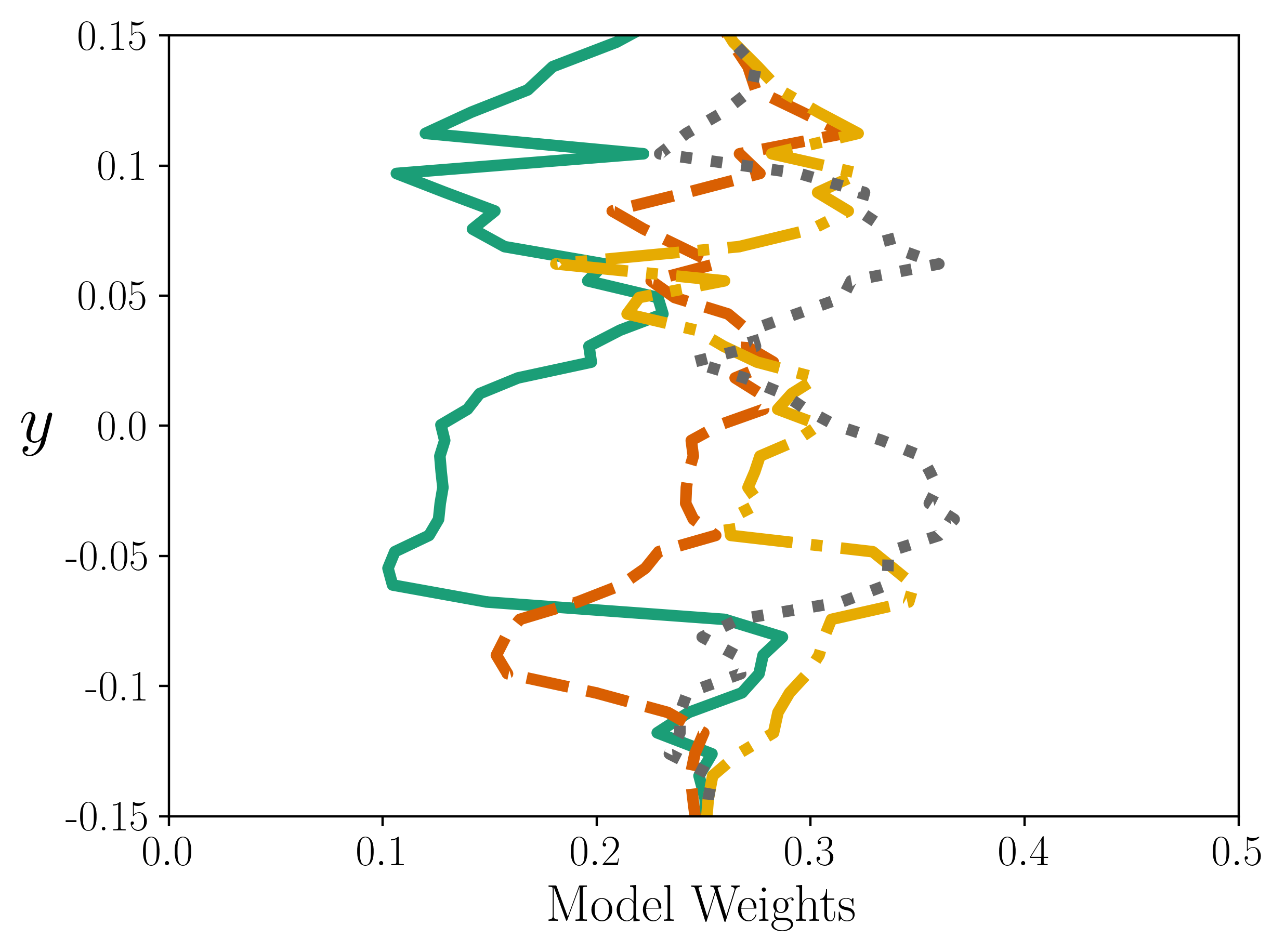} 
    \captionsetup{justification=centering,margin=0cm}
    \caption{Model    weights,    $XMA_1$.}\label{fig_NxN_dim1_Sill1.2_poids_L234_P1}
  \end{subfigure}
  \begin{subfigure}[c]{0.32\linewidth}
    \centering
    \includegraphics[width=\linewidth]
    {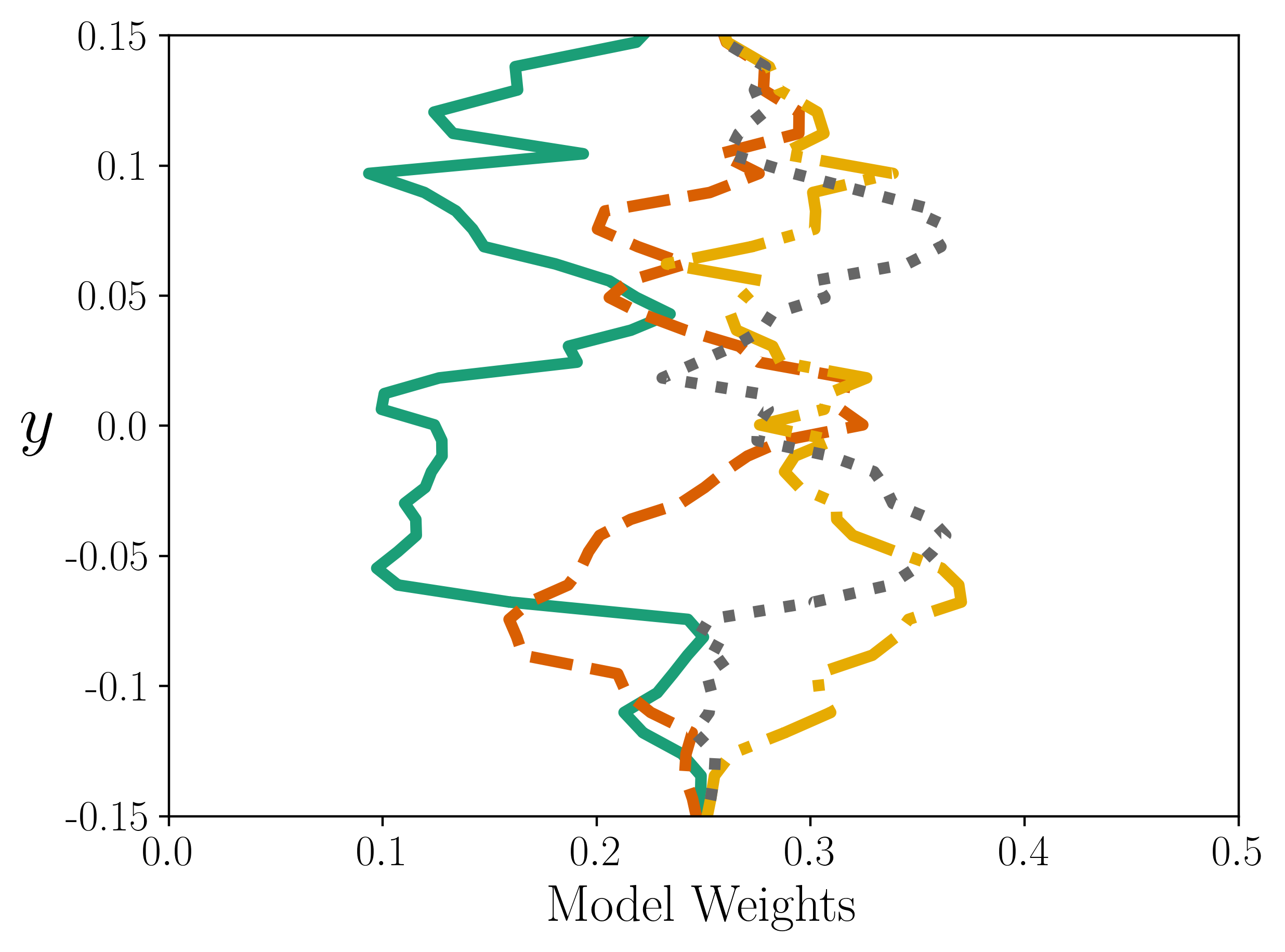} 
        \captionsetup{justification=centering,margin=0cm}
        \caption{Model    weights,    $XMA_2$.}\label{fig_NxN_dim8_Sill1.2_poids_L234_P1}
  \end{subfigure}
  \begin{subfigure}[c]{0.32\linewidth}
    \centering
    \includegraphics[width=\linewidth]
    {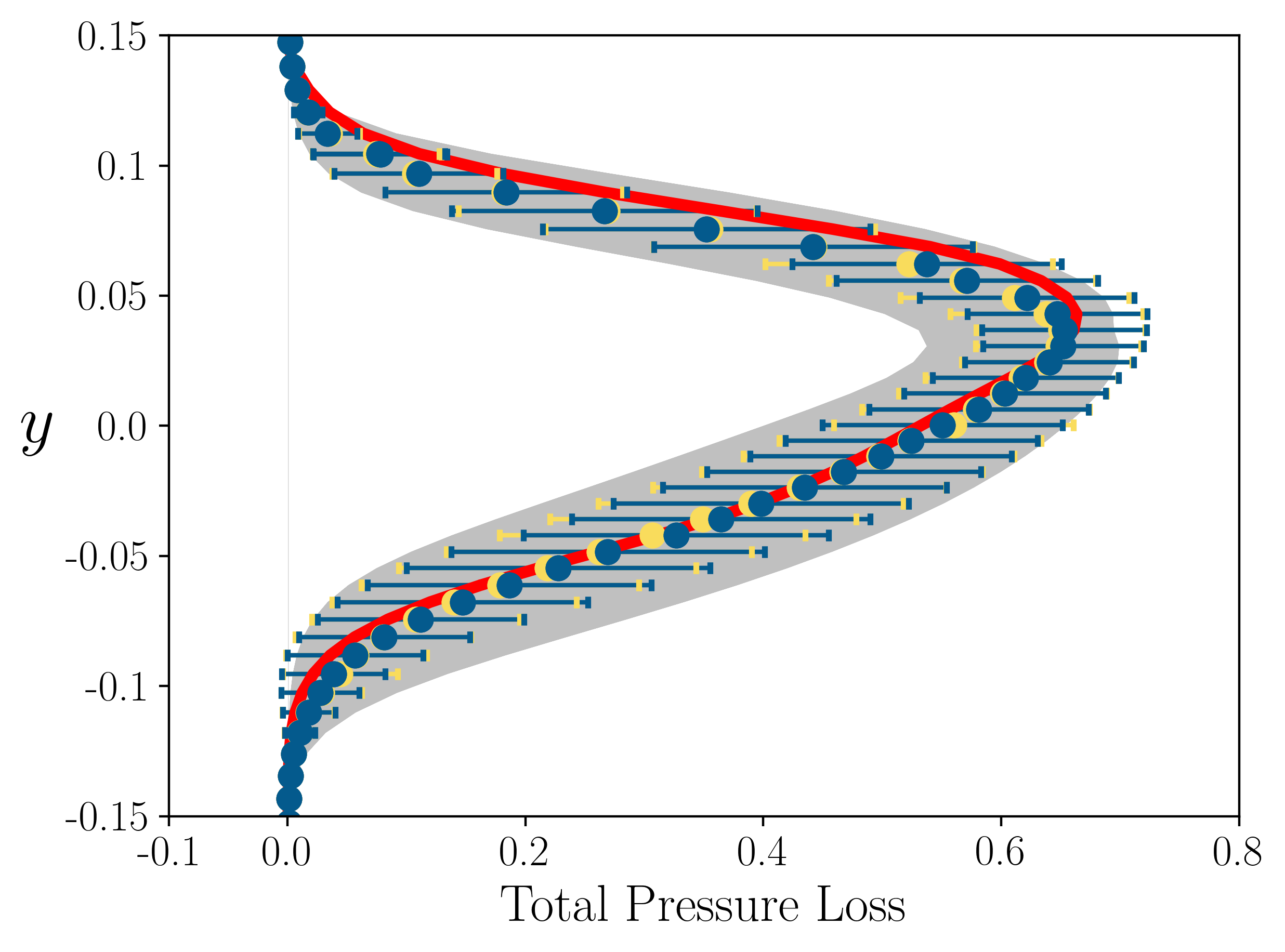}
    \captionsetup{justification=centering,margin=0cm}
    \caption{Prediction}\label{fig_NxN_dim18_Sill1.2_pred_L234_P1}
  \end{subfigure}
  \caption{Profiles of the weighting functions (trained on $\{\Scen_2,\Scen_3,\Scen_4\}$ total
    pressure  data)  and  of  the  total  pressure loss  across  the  wake  at  streamwise  location
    $\frac{x}{l}=1.20$.  Solutions are  reported for  the  big data  ($XMA_1$) and  the scarce  data
    ($XMA_2$) regimes. Prediction on $\Scen_1$.  \\ 
    $\keps$ ($\protect\lsolid{green_Mod}$),  $k-\omega$ model  ($\protect\ldash{orange_Mod}$), $\kl$
    ($\protect\ldashdot{yellow_Mod}$), Spalart-Allmaras ($\protect\ldott{grey_Mod}$), reference data
    ($\protect\lsolid{red}$),  accessible area  \cbox{gray!50}, $\mean{\QOI}  \pm 2\sqrt{\var{\QOI}}$
    for $XMA_1$  ($\protect\bp$), $\mean{\QOI}  \pm \sqrt{\var{\QOI}}$ for  $XMA_2$ ($\protect\yp$).
  }\label{fig_NxN_dim18_Sill1.2_L234_P1}
\end{figure}

%
Fig.   \ref{fig_NxN_dim18_Sill1.2_L234_P1}  presents weighting  functions  and  total pressure  loss
profiles at  $\frac{x}{l}=1.20$ in the wake.   Weighting function profiles for  $XMA_1$ and $XMA_2$,
depicted          in           panels          \ref{fig_NxN_dim1_Sill1.2_poids_L234_P1}          and
\ref{fig_NxN_dim8_Sill1.2_poids_L234_P1}, respectively, are rather  similar, indicating that $XMA_2$
is well informed in this case where we use  three times more data than in the preceding example.  As
observed previously, using more  data leads to sharper model weights.  Due  to the strong similarity
of the  weight distributions,  the total  pressure loss  profiles predicted  by $XMA_1$  and $XMA_2$
(panel \ref{fig_NxN_dim18_Sill1.2_pred_L234_P1})  are very close to  each other.  In both  cases the
reference  data are  captured almost  perfectly,  despite $\Scen_1$  has  not been  included in  the
training  data set.   The reason  for that  is that  the reference  solution is  encompassed by  the
accessible  area,  so that  XMA  can  potentially  fit the  data  if  the weight  distributions  are
accurate. This appears to be the case in  the present example, thus we conclude that XMA generalizes
well to an extrapolation scenario.

In       stark      contrast       with       Fig.       \ref{fig_NxN_dim1_Sill1.2_pred},       Fig.
\ref{fig_NxN_dim18_Sill1.2_pred_L234_P1} presents a large variance on the prediction.  This behavior
can be sourced in two main reasons.  The first reason is that the component RANS models predict very
different solutions for  this configuration, as indicated  by the wide accessible  area.  The second
reason is that the model weights are less sharp on this extrapolation scenario than for an interpolation  scenario, \ie the models are weighted more uniformly.  
Both  features warn  the user  about the  fact that  i) large
model-form uncertainty  exists and ii)  XMA has to  be trusted less  than in the  preceding example.
That being said, the XMA  is shown to be very effective at predicting  QoI for which reference data
lie within the accessible area.  For that purpose,  using a set of component models predicting quite
diverse solutions is beneficial, because it widens the accessible area and increases the probability
of pickling the local best-performing models if the weights are correctly modeled.
%

\begin{figure} 
  \begin{subfigure}[c]{0.32\linewidth}
    \centering
    \includegraphics[width=\linewidth]
    {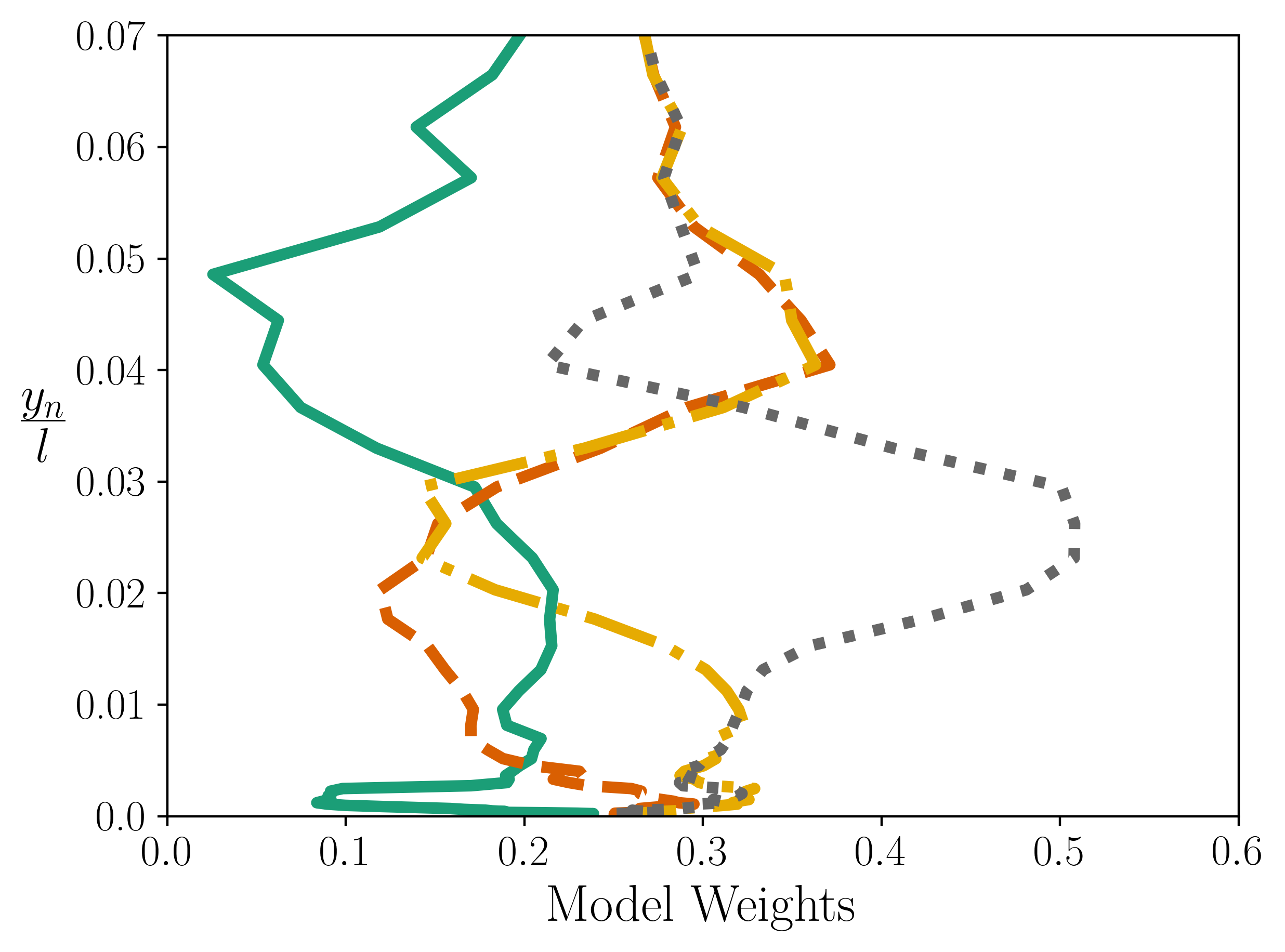}
    \captionsetup{justification=centering,margin=0cm}
    \caption{Model    weights ,    $XMA_1$.}\label{fig_NxN_dim1_Extra0.9_poids_L234_P1}
  \end{subfigure}
  \begin{subfigure}[c]{0.32\linewidth}
    \centering
    \includegraphics[width=\linewidth]
    {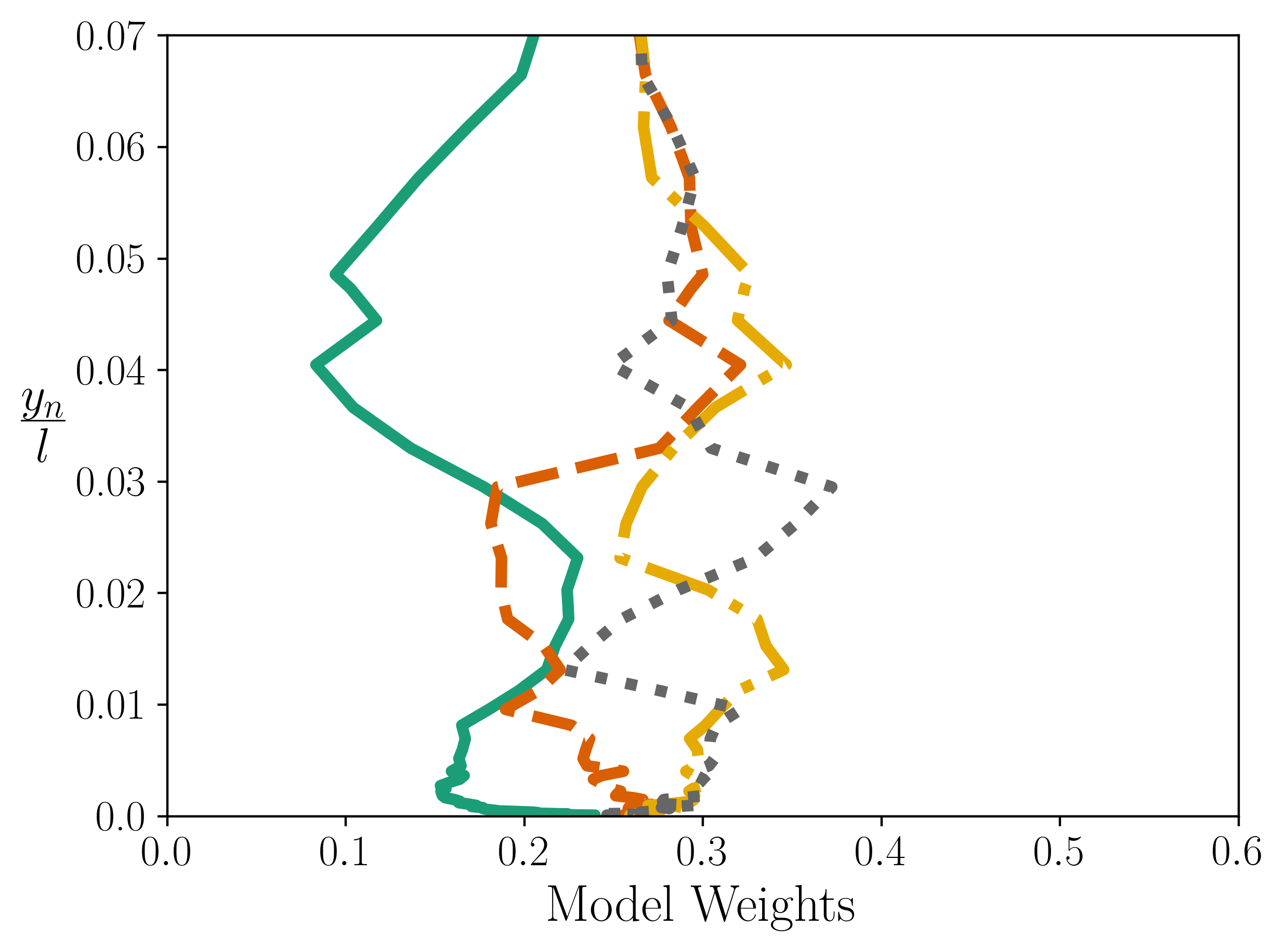}
    \captionsetup{justification=centering,margin=0cm}   
    \caption{Model    weights,    $XMA_2$.}\label{fig_NxN_dim8_Extra0.9_poids_L234_P1}
  \end{subfigure}
  \begin{subfigure}[c]{0.32\linewidth}
    \centering
    \includegraphics[width=\linewidth]
    {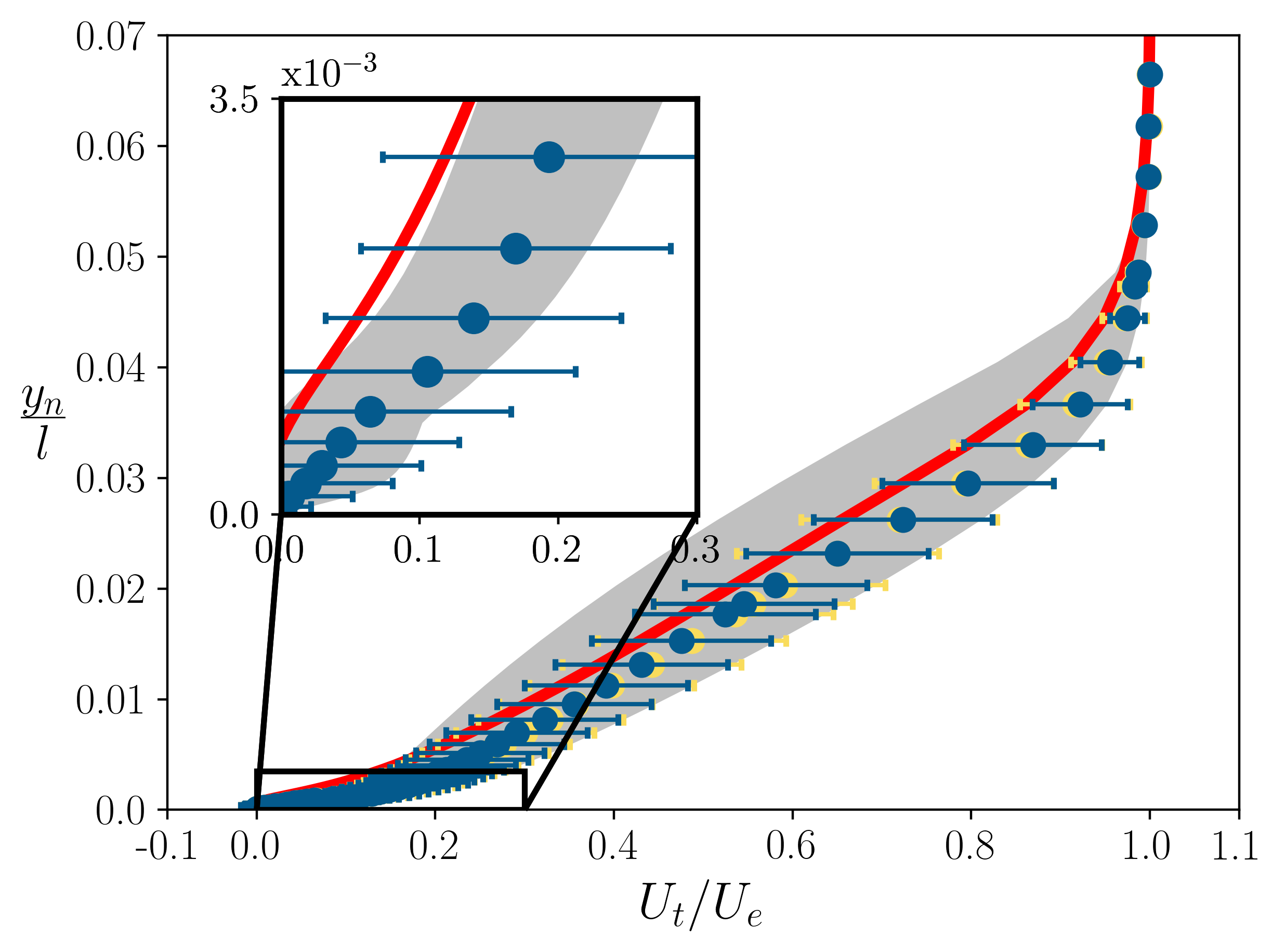}
    \captionsetup{justification=centering,margin=0cm}
    \caption{Prediction}\label{fig_NxN_dim18_Extra0.9_pred_L234_P1}
  \end{subfigure}
  \caption{Profiles  of  the weighting  functions  (trained  on $\{\Scen_2,\Scen_3,\Scen_4\}$  total
    pressure data)  and of the  tangential velocity profile  at streamwise  location $\frac{x}{l}=0.90$.
    Solutions  are   reported  for   the  big   data  ($XMA_1$)  and   the  scarce   data  ($XMA_2$)
    regimes. Prediction on $\Scen_1$.\\ 
    $\keps$   ($\protect\lsolid{green_Mod}$),   $k-\omega$   ($\protect\ldash{orange_Mod}$),   $\kl$
    ($\protect\ldashdot{yellow_Mod}$), Spalart-Allmaras ($\protect\ldott{grey_Mod}$), reference data
    ($\protect\lsolid{red}$), accessible  area \cbox{gray!50}, $\mean{\QOI}  \pm 2\sqrt{\var{\QOI}}$
    ($\protect\bp$).  }\label{fig_NxN_dim1_Extra0.9_L234_P1}
\end{figure}

Fig. \ref{fig_NxN_dim1_Extra0.9_L234_P1} model reports the weight profiles and XMA predictions of the tangential
velocity profile at $\frac{x}{l}=0.90$.
For this quantity not used for training, the solution is less satisfactory, but still in
rather good agreement with the reference data.   The predicted variances are rather large, providing
a measure of RANS modelling uncertainties in the prediction.

\begin{figure} 
  \centering
  \begin{subfigure}[c]{0.50\linewidth}
    \centering
    \includegraphics[width=\linewidth,trim=0 0 0 0, clip]
    {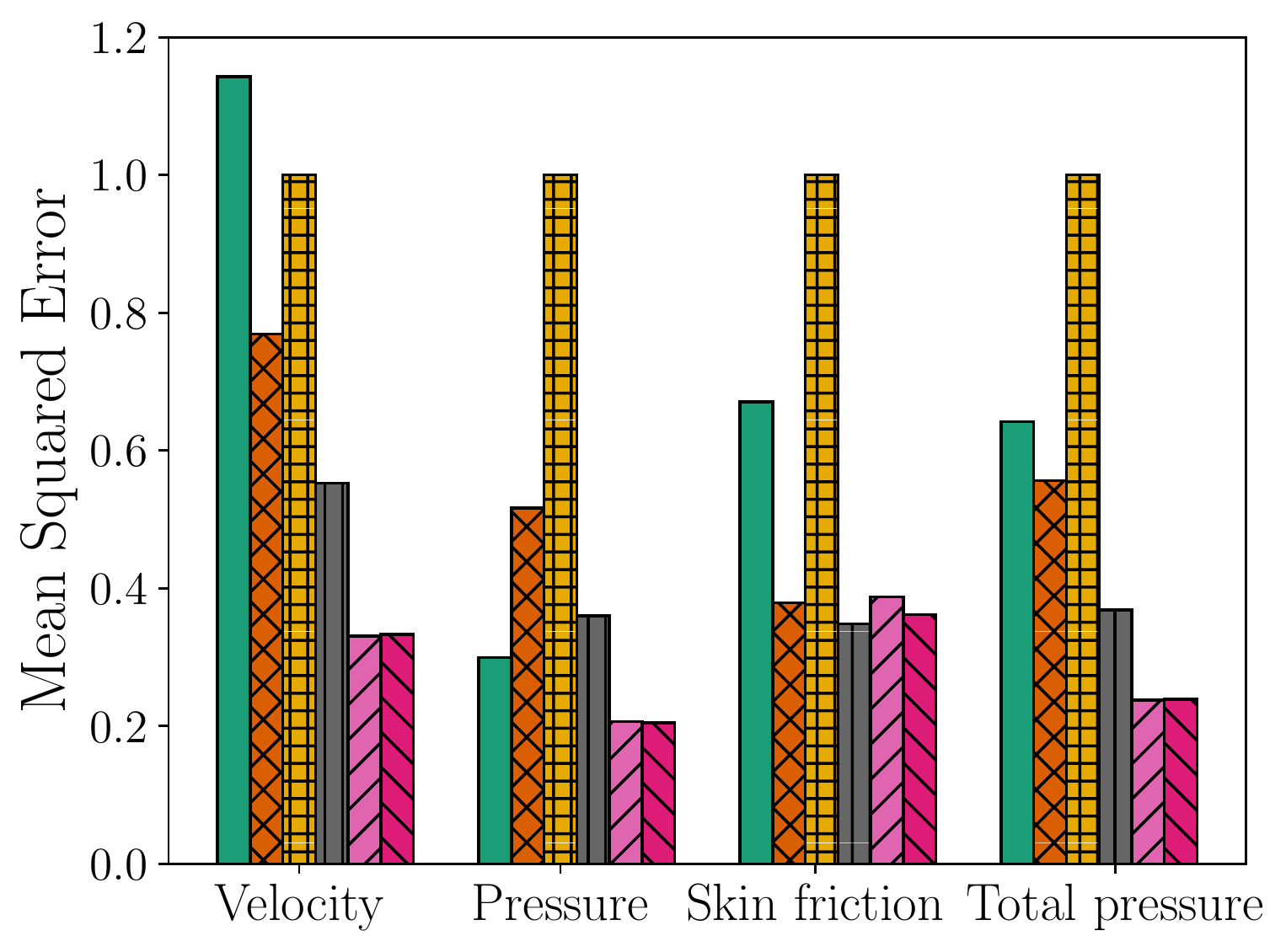} 
  \end{subfigure}
  \caption{ Mean-squared errors for four QoI (normalized by the MSE of the $\kl$ model),
    $\Scen_1$. XMA is trained on $\{\Scen_2,\Scen_3,\Scen_4\}$. \\
    $\keps$  model  ($\protect\bprectKE$),  $k-\omega$  model ($\protect\bprectKO$  ),  $\kl$  model
    ($\protect\bprectKL$)      and     Spalart-Allmaras      model     ($\protect\bprectSA$), $XMA_1$
    ($\protect\bprectXBMAA$), $XMA_2$ ($\protect\bprectXBMAB$). }\label{fig_NxN_MSE_adimKE_L234_P1}
\end{figure}

Finally, in Fig.  \ref{fig_NxN_MSE_adimKE_L234_P1} we report the MSE for various QoI.  In this case
the results are  normalized with the error of  the $\kl$ model, which exhibits the  largest error on
$3$ out of $4$ QoI. 

The XMA predictions clearly  improve the accuracy over the component RANS  models, regardless of the
QoI presented.  For example, the MSE for the velocity is reduced by approximately $1/3$ with respect
to the best-performing baseline RANS model.  Here again,  $XMA_1$ and $XMA_2$ lead to similar MSE on
average, showing that the scarce data regime is already sufficient to properly inform the mixture.

\section{Conclusions}\label{chapXBMA_conclusion}
A novel space-dependent  Model Aggregation (XMA) algorithm is introduced  and assessed for improving
Reynolds-Averaged predictions  of turbulent flows  while providing estimates of  turbulence modeling
uncertainties.  XMA is a  multi-model ensemble technique that builds a  convex linear combination of
RANS solutions  obtained by using  a set of  competing turbulence models,  whose weight depend  on a
vector of well-chosen local flow features. A supervised machine learning algorithm (Random
Forests) is used to regress  the mixture weights as functions of the features  from a training set of
observed  flow quantities.   The  weighting functions  are  based on  a cost  function  that can  be
interpreted  as  the  likelihood  for  a  given  model to  capture  the  observed  data,  given  the
features. They  can be interpreted  as a  relative score of  performance assigned to  each component
model in the set of candidates.  A convex  linear combination of the individual RANS model solutions
and the  weighting functions is  used to estimate  an average RANS  solution and its  variance.  The
latter provides a measure  of the uncertainty in the prediction, due to  the lack of knowledge about
the best-performing  turbulence model for  a given  flow scenario and  local flow features.   In the
numerical examples, we  focus on a set of  four linear eddy viscosity models (LEVM),  widely used in
engineering applications. Specifically, the  Spalart-Allmaras, Launder-Sharma $\keps$, Wilcox (2006)
$k-\omega$, and Smith $k-L$  models are used to build the XMA mixture.  The  approach is assessed for
flow configurations of practical interest, namely, compressible turbulent flows through the NACA $65$
V$103$ planar compressor  cascade.  For proving the interest of  the proposed methodology, synthetic
data  are generated  from RANS  simulations based  on an  explicit algebraic  Reynolds stress  model
(EARSM), i.e. a model with a significantly different turbulence anisotropy structure compared to the
LEVM.  The EARSM reference simulations, obtained for a set of four different operating conditions of
the cascade (called "scenarios"),  provides target values for all possible  flow quantities for both
training and validation.  The synthetic data are  used to investigate the influence of observed flow
quantities  and of  the  size of  the training  set  on the  quality  of XMA  predictions.  For  the
configuration  at stake,  the total  pressure,  which depends  on both  kinematic and  thermodynamic
quantities,  is used to train the XMA weights.  Various training
datasets  are constructed  by uniformly  sampling  the reference  EARSM solutions.   The latter  may
contain observations extracted  from a single flow scenario or  by multiple scenarios simultaneously
and both "big data" and "small data" regimes are considered.  The first one corresponds to including
total pressure  data at  all points  of the computational  mesh for  the reference  solution(s). The
second one  corresponds to extracting information  at one mesh point over 8, leading to  a dataset of
only $820$ data if a single scenario is included.

When XMA is trained on a single scenario and used to predict the same scenario, the predicted fields
of  the  observed quantity  do  not  match  the reference  data  perfectly,  due to  the  structural
deficiencies of the underlying LEVM, whose individual  solutions do not encompass the reference once
everywhere in the flow.  However, on the average, XMA consistently improves the  results over any of
the component models, because it consistently assigns  higher weighting to the best performing models
in each flow region.  Interestingly, XMA does not only improve the prediction  for the flow quantity
used for training,  but also for unobserved  quantities that are reasonably well correlated with the preceding one, such as the  velocity field.  Additionally,
the learned weighting functions and component solutions can  be used to estimate the variance of the
XMA estimate, which represents a measure of the  consensus among the candidate models about what the
solution  should be.   Relatively similar  results are  obtained  for both  for the  big and  scarce
training  sets, with  the accuracy  increasing and  the  variance decreasing  with the  size of  the
training set.

Afterward, XMA  is trained  against three flow  scenarios and  used to predict  a fourth  one, whose
operating conditions are not within the range  of operating conditions of the calibration scenarios.
For such  extrapolation situation,  the XMA  shows very good  generalization abilities,  providing a
solution that is overall more accurate than any  of the component models for most flow quantities of
interest.  In this case  however, the variances are larger, warning the  user about the trustfulness
of the results.

The present XMA relies  on "on the shelf" component turbulence  models, not specifically calibrated
for the configuration of  interest. Additionally, no attempt was done to  optimize the placement of
observation  points  used  for  model  training. Thus,  possible  future  improvements  consist  in
simultaneously calibrating  the models and  training the  weighting functions, as well as  utilizing optimal
sensor placement  techniques (e.g. \cite{mons2017optimal}) for  selecting the training data.  Further work on
the choice of the flow features used to describe the model weights is also planned.

\section{Acknowledgement}
The present  work was co-funded  by Safran  Tech and ANRT  (Agence Nationale de  la Recherche
  Technologique)  under CIFRE  Grant N.  N2018/1370. The  authors also  thank CMasher  for providing  the
  colormaps used in the present plots.


\bibliographystyle{unsrt}
\bibliography{biblio.bib}

\end{document}